\documentclass[twocolumn,superscriptaddress,showkeys,nobibnotes]{revtex4}
\usepackage{amsfonts}
\usepackage{amssymb}
\usepackage{amsmath}
\usepackage{epsfig}
\usepackage{color}
\usepackage{graphics, graphicx}
\usepackage{bbold}
\usepackage{psfrag}
\usepackage{mathcomp}
\usepackage{subfigure}
\usepackage{verbatim}
\usepackage{float}
\usepackage{graphicx}
\usepackage[colorlinks,citecolor=blue]{hyperref}
\makeatletter

\newcommand{\Rmnum}[1]{\expandafter\@slowromancap\romannumeral #1@}
\makeatother
\bibpunct{[}{]}{,}{n}{}{,}

\begin{document}

\title{Rydberg-Atom-Mediated Strong Antisymmetric Spin Exchange in Molecular Arrays}
%\title{Rydberg-atom-mediated strong  spin-exchange interactions of ultracold molecules in optical tweezer arrays}

\author{Yunqing Jiao}
\thanks{These authors contributed equally to this work.}
\affiliation{State Key Laboratory of Quantum Optics Technologies and Devices, Institute of Laser Spectroscopy, Shanxi University, Taiyuan, Shanxi 030006, China}

\author{Jin-Zhu Jiang}
\thanks{These authors contributed equally to this work.}
\affiliation{State Key Laboratory of Quantum Optics Technologies and Devices, Institute of Laser Spectroscopy, Shanxi University, Taiyuan, Shanxi 030006, China}

\author{Bo-Wen Guan}
\affiliation{State Key Laboratory of Quantum Optics Technologies and Devices, Institute of Laser Spectroscopy, Shanxi University, Taiyuan, Shanxi 030006, China}

\author{Jie Ma}
\affiliation{State Key Laboratory of Quantum Optics Technologies and Devices, Institute of Laser Spectroscopy, Shanxi University, Taiyuan, Shanxi 030006, China}
\affiliation{Collaborative Innovation Center of Extreme Optics, Shanxi University, Taiyuan, Shanxi 030006, China}

\author{Liantuan Xiao}
\affiliation{State Key Laboratory of Quantum Optics Technologies and Devices, Institute of Laser Spectroscopy, Shanxi University, Taiyuan, Shanxi 030006, China}
\affiliation{Collaborative Innovation Center of Extreme Optics, Shanxi University, Taiyuan, Shanxi 030006, China}

\author{Chi Zhang}
\email{c.zhang@imperial.ac.uk}
\affiliation{Centre for Cold Matter, Blackett Laboratory, Imperial College London, Prince Consort Road, London SW7 2AZ, United Kingdom}

\author{Weibin Li}
\email{weibin.li@nottingham.ac.uk}
\affiliation{School of Physics and Astronomy, and Centre for the Mathematics and Theoretical Physics of Quantum Non-equilibrium Systems, The University of Nottingham, Nottingham NG7 2RD, United Kingdom}

\author{Feng Mei}
\email{meifeng@sxu.edu.cn}
\affiliation{State Key Laboratory of Quantum Optics Technologies and Devices, Institute of Laser Spectroscopy, Shanxi University, Taiyuan, Shanxi 030006, China}
\affiliation{Collaborative Innovation Center of Extreme Optics, Shanxi University, Taiyuan, Shanxi 030006, China}

\author{Suotang Jia}
\affiliation{State Key Laboratory of Quantum Optics Technologies and Devices, Institute of Laser Spectroscopy, Shanxi University, Taiyuan, Shanxi 030006, China}
\affiliation{Collaborative Innovation Center of Extreme Optics, Shanxi University, Taiyuan, Shanxi 030006, China}

\begin{abstract}
	Ultracold molecular systems have recently emerged as a versatile platform for quantum computation and simulation. Spin-exchange interactions arising from direct molecular dipolar interactions constitute the key mechanism for generating quantum entanglement and simulating quantum spin models. However, the relatively small electric dipole moments result in weak spin-exchange couplings, fundamentally limiting the speed of quantum information processing and interaction cycle of many-body dynamics. Here, we introduce a framework that employs Rydberg atoms with large electric dipole moments to mediate strong interactions between molecules in optical tweezer arrays that enables individually laser addressing both the Rydberg atom and molecules. Our result reveals that the mediated coupling can realize an effective molecular spin-exchange interaction with an intrinsic Dzyaloshinskii-Moriya character, with the Rydberg atoms dynamically decoupled from the molecular degrees of freedom, and the effective interaction strength enhanced by up to three orders of magnitude. We further demonstrate its versatility through rapid entanglement generation, high-fidelity two-qubit gate operations, and the realization of non-equilibrium symmetry-protected topological phase with long-lived edge coherence. Our work establishes a route toward strong molecular spin interactions and opens opportunities for fast, scalable quantum information processing and quantum simulation of long-time non-equilibrium quantum many-body physics in optical tweezer arrays of ultracold molecules.
\end{abstract}


\maketitle

\section*{1.~Introduction}

Recent advances in ultracold molecules have positioned them at the forefront of quantum science owing to their rich internal structure, long coherence times, and long-range dipolar interactions~\cite{NPreview_Maniandcooling_24}, which enable a broad range of applications, including quantum computation and simulation~\cite{NPreview_QCQS_24}, precision metrology~\cite{NPreview_Metrology_24}, and ultracold chemistry~\cite{NPreview_Chemistry_24}. In particular, the rapid development of optical tweezer technologies and microwave shielding techniques~\cite{Shield_MSPM_prl18,Shield_Scatter_prl18,Shield_Doyle_21Sci,Shield_IBloch_22Nat,Shield_TaoS_23L,Shield_field-linked_N23,Shield_tetratomic_N24,Shield_TaoS_25L,Shield_XCui_26L,Shield_TaoS_26NP} has enabled high-fidelity preparation~\cite{tweezer_19,tweezer_20,tweezer_Zhan_20,tweezer_ployatomic_24}, cooling~\cite{tweezer_preparation_22,tweezer_raman_24,tweezer_raman3D_24,tweezer_3DMOT_BYprl24} and coherent control~\cite{tweezer_JingZ_14,tweezer_storage_21,tweezer_enhanced_24,tweezer_site-selective_24,tweezer_spin1_25,entangle_squeezing_26,tweezer_MBSD_Cheuk26} of ultracold molecules. Building on these capabilities, recent experiments have for the first time reported the generation of quantum entanglement between ultracold molecules in optical tweezer arrays~\cite{entangle_science_bao_23,entangle_science_Holland_23,entangle_nature_iSWAP_25,entangle_nature_longlived_25}, leveraging spin-exchange interactions induced by molecular dipolar interactions and opening the prospect for ultracold molecular quantum computing~\cite{entangle_PRL02_DeMille,entangle_KKNi_CS18,NPreview_QCQS_24}. However, the dipolar interaction strengths remain fundamentally limited by the relatively small transition dipole moments, which are typically on the order of a few Debye. As a result, dipolar spin-exchange interactions generally lie in the tens-of-hertz regime at micrometer separations, leading to entanglement-generation times on the millisecond scale~\cite{entangle_science_bao_23,entangle_science_Holland_23}. 

In parallel, Rydberg atoms possess giant transition dipole moments, typically several orders of magnitude larger than those of polar molecules, giving rise to strong interactions in the hundreds of megahertz or gigahertz regime and making them a powerful entangling resource for quantum information processing~\cite{Rydberg_review_20,Rydberg_review_21}. The combination of strong Rydberg interactions with the long coherence times and rich internal structure of ultracold molecules has motivated hybrid molecule-atom architectures~\cite{hybrid_PRXquantum_ChiZhang_22,hybrid_PRXquantum_enriching_22,hybrid_GHZ_26}, which have subsequently been explored experimentally~\cite{hybrid_blockade_23,hybrid_probing_25,hybrid_Harness_26}, opening a promising route toward extending the capabilities of molecular quantum platforms~\cite{hybrid_readout_Yelinpra16,hybrid_readout_26,hybrid_CNOT_26}. However, generating Rydberg-atom-mediated spin-exchange interaction Hamiltonian solely between molecules remains an unresolved challenge, which is essential for advancing both molecular quantum computation and quantum simulation of spin chain models.

In this work, we present a hybrid molecule–atom framework in which a Rydberg atom serves as a quantum bus mediating strong spin interactions between molecules. We consider an optical array setting, where both the Rydberg atom and molecules are laser addressed individually. Through a two-step Rydberg-mediated protocol, the ancillary Rydberg degree of freedom is dynamically decoupled, thereby producing an effective molecular spin Hamiltonian that realizes a strong antisymmetric spin-exchange interaction between molecules with an intrinsic Dzyaloshinskii-Moriya (DM) character. We first use this interaction to demonstrate rapid entanglement generation and high-fidelity two-qubit gate operations under realistic experimental conditions. Compared with existing molecule-only implementations, the protocol achieves operation times up to three orders of magnitude shorter while remaining robust against dominant decoherence mechanisms. Beyond serving as a fast two-qubit resource, we then exploit the DM interaction to efficiently prepare one- and two-dimensional cluster states for implementing large-scale measurement-based molecular quantum computation. Finally, we use this interaction to realize a non-equilibrium symmetry-protected topological phase. Through complementary static and dynamical diagnostics, we identify the characteristic signatures of the phase, including long-lived edge coherence. Together, these results establish the hybrid molecule–atom framework as a platform for ultracold molecular quantum computation and simulation.

\section*{2.~Results}
\subsection*{Rydberg-atom-mediated strong spin-exchange interaction between molecules}\label{Model}

\begin{figure*}[htbp]
	\centering
	\includegraphics[scale=0.72]{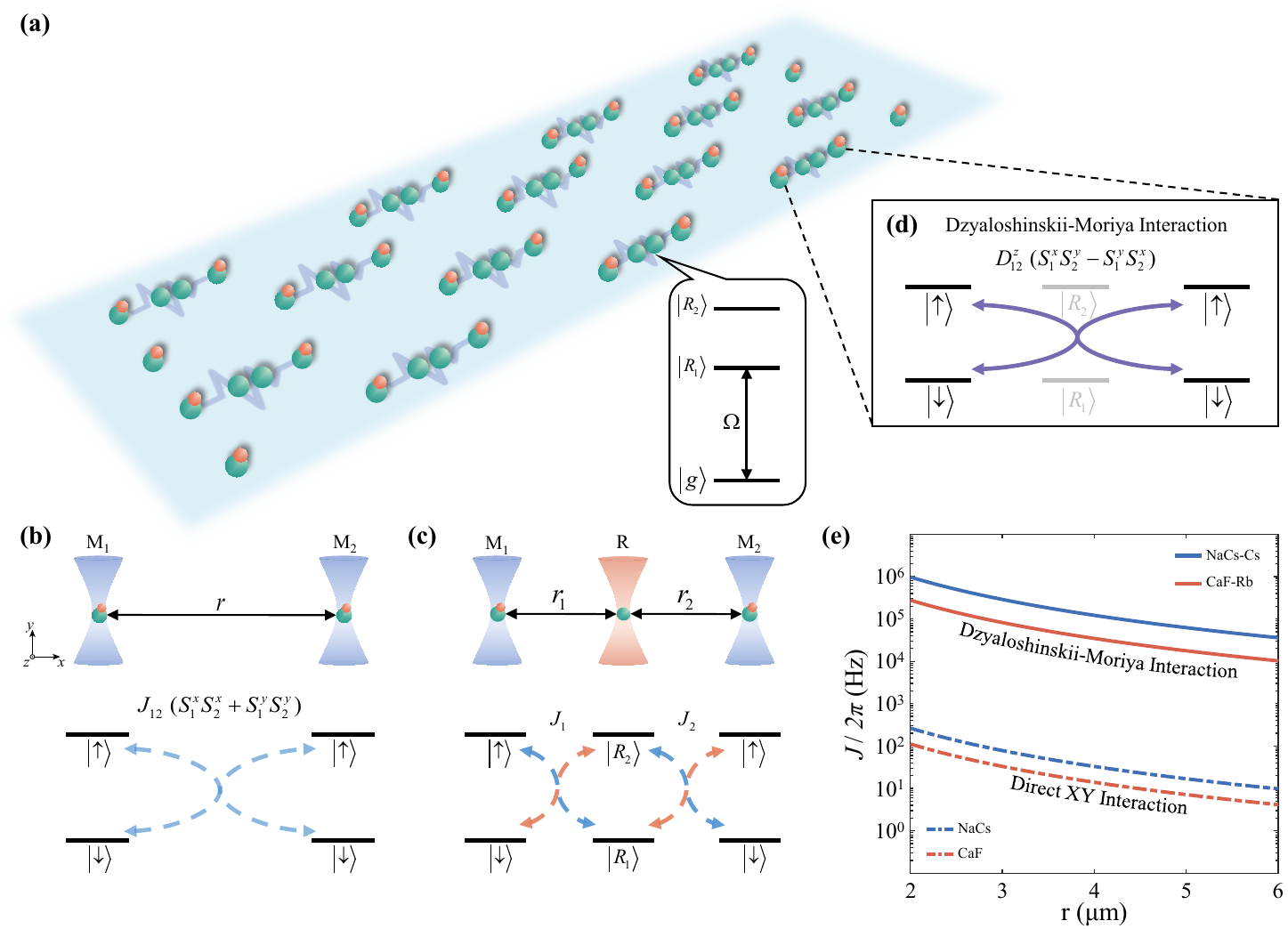}
	\caption{\textbf{Rydberg-atom-mediated strong spin-exchange interaction between molecules.} 
    (a) Schematic illustration of a hybrid optical tweezer array of ultracold molecules and Rydberg atoms.
    (b) Dipole-dipole interaction between ultracold molecules (top) with its resulting XY spin-exchange interaction (bottom). 
    (c) Dipole-dipole interaction between Rydberg atoms and ultracold molecules (top) and the corresponding spin-exchange interaction (bottom), with the intermolecular distance being $r=r_1+r_2$.
    (d) Rydberg-atom-mediated DM spin-exchange interaction between ultracold molecules. 
    (e) Spin-exchange strengths of the DM interaction (solid line) and the XY interaction (dashed line) as a function of distance. The molecule-only results correspond to NaCs and CaF, while the hybrid results use the NaCs--Cs and CaF--Rb combinations, respectively. The transition dipole moments are $d^{\uparrow\downarrow}_{\mathrm{NaCs}} = 2.644\ \mathrm{Debye}$, $d^{\uparrow\downarrow}_{\mathrm{CaF}} = 1.72\ \mathrm{Debye}$, $d^{\uparrow\downarrow}_{\mathrm{Cs}} = 11220\ \mathrm{Debye}$ and $d^{\uparrow\downarrow}_{\mathrm{Rb}} = 4911\ \mathrm{Debye}$.}
	\label{Fig1}
\end{figure*}
               
We consider a hybrid optical tweezer array consisting of polar molecules and auxiliary Rydberg atoms, as illustrated in Fig.~\ref{Fig1}a. Molecular pseudospins are encoded in long-lived rotational states, while the Rydberg atoms serve as controllable mediators positioned between neighboring molecules. The goal of this framework is to realize an effective spin-exchange interaction Hamiltonian between molecules while leveraging the strong dipolar interactions available in Rydberg systems. The motivation for introducing the auxiliary atoms becomes apparent when compared with molecule-only frameworks, where interactions are generated through direct dipolar spin exchange between molecular rotational states~\cite{DDI_BoYanNature_13}, as schematically illustrated in Fig.~\ref{Fig1}b. However, because these interactions are set primarily by molecular transition dipole moments, their strengths are typically limited to the tens-of-hertz range at micrometer separations commonly used in molecular-tweezer experiments, leading to comparatively slow entangling operations.

Within the hybrid array, the Rydberg-atom-mediated spin-exchange interaction is illustrated in Fig.~\ref{Fig1}c. Its underlying mechanism is captured by an elementary three-body subsystem consisting of two ultracold polar molecules ($\mathrm{M}_1,\mathrm{M}_2$) and an auxiliary Rydberg atom ($\mathrm{R}$). The relevant molecular and Rydberg transitions are taken to be resonant, such that resonant spin exchange dominates the dynamics. The microscopic molecule-atom dipole-dipole Hamiltonian is
\begin{equation}
	H_{\mathrm{dd}} = \sum_{n=1,2} \frac{\kappa}{r_{n}^3} \left[
	\mathbf{d}_{n} \cdot \mathbf{d}_{\mathrm{R}}
	- 3(\mathbf{d}_{n} \cdot \mathbf{\hat{r}}_{n})
	(\mathbf{d}_{\mathrm{R}} \cdot \mathbf{\hat{r}}_{n})
	\right],
    \label{Hdd}
\end{equation}
where $\kappa = 1/(4\pi\varepsilon_0\hbar)$ with $\hbar=1$, $\mathbf{d}_{n}$ and $\mathbf{d}_{\mathrm{R}}$ denote the electric dipole moment operators of molecule $n$ and the Rydberg atom respectively. $r_n$ is the molecule-atom separation and $\hat{\mathbf r}_n$ is the corresponding unit vector. Projecting onto the effective spin-1/2 pseudospin manifold $\{|{\downarrow}\rangle, |{\uparrow}\rangle\}_{\textrm{M}_1} \otimes \{|{R_1}\rangle, |{R_2}\rangle\}\otimes\{|{\downarrow}\rangle, |{\uparrow}\rangle\}_{\textrm{M}_2}$, yields the effective spin-exchange Hamiltonian (see Methods):
\begin{equation}
	H = \frac{1}{2} \left[
	J_1 (S_{1}^{+} S_{\mathrm{R}}^{-} + \mathrm{H.c.})
	+ J_2 (S_{\mathrm{R}}^{+} S_{2}^{-} + \mathrm{H.c.})
	\right],
	\label{H}
 \end{equation}
where $S_{n}^{+}$ and $S_{\mathrm{R}}^{+}$ denote Pauli raising operators  of molecule n and the Rydberg atom respectively, with $S^+=(S^x+iS^y)/2$. The distance-dependent interaction strengths are determined by the transition dipole moments
\begin{equation}
	J_{n} = \frac{{\kappa  d^{\uparrow\downarrow}_{n}}  d^{\uparrow\downarrow}_{\mathrm{R}}}{r^{3}_{n}},
	\label{J}
\end{equation}
where $d^{\uparrow\downarrow}_{n}=\langle\uparrow|\mathbf d_{n}|\downarrow\rangle$ and 
$d^{\uparrow\downarrow}_{\mathrm R}=\langle R_2|\mathbf d_{\mathrm R}|R_1\rangle$ denote the transition dipole moments of the molecules and the Rydberg atom respectively, evaluated within the pseudospin manifolds. The interaction strengths are orientation-independent since we set the quantization axis $z$ perpendicular to the molecular plane $x-y$ (Fig.~\ref{Fig1}b inset). This Hamiltonian captures the dominant coherent spin-exchange processes mediating population transfer between the molecular pseudospins via the Rydberg atom.

To present the effective spin-exchange interaction Hamiltonian acting on molecular pseudospins through Rydberg-mediated dynamics, we next resolve the Hamiltonian Eq.~\eqref{H} according to the conserved total excitation number. Restricted to the pseudospin manifolds defined above, the hybrid system comprises eight basis states, which naturally separate into independent dipolar excitation sectors:
$n_{\textrm{exc}}=0:\{ \left| \downarrow R_1 \downarrow \right.\rangle \},$ 
$n_{\textrm{exc}}=1:\{ \left| \downarrow R_1 \uparrow \right.\rangle,  \left| \uparrow R_1 \downarrow \right.\rangle,  \left| \downarrow R_2 \downarrow \right.\rangle \},$ 
$n_{\textrm{exc}}=2:\{ \left| \downarrow R_2 \uparrow \right.\rangle,  \left| \uparrow R_2 \downarrow \right.\rangle,  \left| \uparrow R_1 \uparrow \right.\rangle \},$ 
$n_{\textrm{exc}}=3:\{ \left| \uparrow R_2 \uparrow \right.\rangle \}$. 
Dipolar interactions couple states exclusively within each excitation manifold, rendering dynamics across different sectors independent. The sector Hamiltonians are
\begin{equation}
	\begin{aligned}
		H_{n_{\textrm{exc}}=1} &= \frac{1}{2} \begin{pmatrix}
		0 & 0 & J_2 \\
		0 & 0 & J_1 \\
		J_2 & J_1 & 0
	\end{pmatrix},\\
	H_{n_{\textrm{exc}}=2} &= \frac{1}{2} \begin{pmatrix}
		0 & 0 & J_1 \\
		0 & 0 & J_2 \\
		J_1 & J_2 & 0
	\end{pmatrix}.
	\end{aligned}
    \label{Hn}
\end{equation}

Motivated by the structure of these sector Hamiltonians, we construct a two-step protocol in which two Rydberg atoms are placed at independently chosen positions to generate a strong effective spin-exchange interaction Hamiltonian between the molecules (see Fig.~\ref{Fig1}a). Specifically, the position of atom 1 is chosen such that $J_{2\alpha}/{\tilde J_\alpha} = \sin(\alpha/2)$ and $J_{1\alpha}/{\tilde J_\alpha} = \cos(\alpha/2)$, where $ \tilde J_\alpha = \sqrt{J_{1\alpha}^2 + J_{2\alpha}^2} $. Combining these relations with Eq.~\eqref{J} yields $\tan(\alpha/2) = J_{2\alpha}/J_{1\alpha} = r_{1\alpha}^3/r_{2\alpha}^3$. Atom 1 is first initialized to the Rydberg state $ |R_1 \rangle $, and the system then undergoes free evolution for a time interval $ t_1 = 2\pi /\tilde J_\alpha $, which sets the characteristic timescale of the Rydberg-mediated dynamics. At the end of this interval, the Rydberg atom is flipped back to the ground state $|g\rangle$, removing it from the interacting Rydberg manifold and thereby switching off its dipolar coupling to the molecules. According to Eq.~\eqref{J}, owing to the giant transition dipole matrix elements between Rydberg states, the hybrid spin-exchange interaction is several orders of magnitude stronger than the molecule-molecule spin-exchange interaction. Therefore, on the timescale $t_1$, direct dipolar exchange between molecules can be safely neglected and the dynamics is governed entirely by the Rydberg-mediated interactions. The resulting sector evolution operators $U_{n_{\mathrm{exc}}}=\exp(-i H_{n_{\mathrm{exc}}} t_1)$ are
\begin{equation}
	\begin{aligned}
		U_{n_{\textrm{exc}}=1} &= \begin{pmatrix}
		\cos\alpha & -\sin\alpha & 0 \\
		-\sin\alpha & -\cos\alpha & 0 \\
		0 & 0 & -1
	\end{pmatrix}, \\
		U_{n_{\textrm{exc}}=2} &= \begin{pmatrix}
		-\cos\alpha & -\sin\alpha & 0 \\
		-\sin\alpha & \cos\alpha & 0 \\
		0 & 0 & -1
	\end{pmatrix}.
	\end{aligned}
	\label{Un}
\end{equation}
To make the effective evolution induced between the two molecules explicit, we express the unitary dynamics in the basis $\{\left|\downarrow R_1\downarrow\right.\rangle, \left|\downarrow R_1\uparrow\right.\rangle, \left|\uparrow R_1\downarrow\right.\rangle, \left|\uparrow R_1\uparrow\right.\rangle\}$, in which the Rydberg atom occupies a fixed reference state. In this representation, the unitary evolution generated during the first free-evolution segment takes the block-diagonal form:
\begin{align}
	U_1(\alpha) = \begin{pmatrix}
	1 & 0 & 0 & 0 \\
	0 & \cos\alpha & -\sin\alpha & 0 \\
	0 & -\sin\alpha & -\cos\alpha & 0 \\
	0 & 0 & 0 & -1
	\end{pmatrix}.
	\label{U1}
\end{align}
Analogously, atom 2 is positioned to satisfy
$J_{2\beta}/{\tilde J_\beta} = \sin(\beta/2)$, $J_{1\beta}/{\tilde J_\beta} = \cos(\beta/2)$ with $ \tan(\beta/2) = J_{2\beta}/J_{1\beta} = r_{1\beta}^3/r_{2\beta}^3 $. Following initialization of atom~2 in $|R_1\rangle$, the system evolves freely for $t_2 = 2\pi /\tilde J_\beta$, after which the atom is driven back to $|g\rangle$. The composite unitary operation $ U_2 = U_1(\beta)U_1(\alpha) $ is implemented after a total time $ t = t_1 + t_2 $:
\begin{align}
	U_{2}(\alpha-\beta)= \begin{pmatrix}
	1&0&0&0\\
	0&\cos(\alpha-\beta) & - \sin (\alpha-\beta) &0\\
	0&\sin (\alpha-\beta) &\cos (\alpha-\beta) &0\\
	0&0&0&1
	\end{pmatrix}.
	\label{U2}
\end{align}

In the basis chosen above, the states $\left|\downarrow R_1 \downarrow\right\rangle$ and $\left|\uparrow R_1 \uparrow\right\rangle$, which belong to different excitation sectors, are dynamically decoupled. We therefore focus on the $n_{\mathrm{exc}} = 1$ subspace $\{\left|\downarrow R_1 \uparrow\right\rangle,\, \left|\uparrow R_1 \downarrow\right\rangle\}$. As shown, although the intermediate Rydberg state $|R_2\rangle$ is involved during the dynamics~(Eq.~\eqref{Hn}), the atoms return to $|R_1\rangle$ at the end of each evolution cycle~(Eq.~\eqref{Un}). Accordingly, the atomic degree of freedom is decoupled from the molecular pseudospin manifold, and the effective evolution can be described solely in the molecular subspace. The molecular evolution can be further written as $\exp(-iH_{\mathrm{DM}}^z t)$, with
\begin{equation}
    H_{\mathrm{DM}}^z=D_{12}^z(\mathbf S_1\times\mathbf S_2)_z=D_{12}^z(S_1^xS_2^y-S_1^yS_2^x),
    \label{HDM}
\end{equation}
where $D_{12}^z=(\alpha-\beta)/2t$ is the antisymmetric DM spin-exchange interaction strength (see Supplementary Note~1 for detailed derivation). This Hamiltonian corresponds to the $z$-component of the DM interaction, as illustrated in Fig.~\ref{Fig1}d.

To quantify the interaction enhancement enabled by the Rydberg mediator, we compare the strengths of the direct molecular XY interaction and the Rydberg-mediated effective DM interaction as functions of intermolecular distance. Recent experiments have reported dipolar spin-exchange entanglement in optical tweezer arrays of CaF and NaCs molecules~\cite{entangle_science_bao_23,entangle_science_Holland_23,entangle_nature_iSWAP_25}, motivating these two molecular species as representative choices. In our calculations, NaCs and CaF are considered for the molecule-only scheme, while NaCs--Cs and CaF--Rb are used for the corresponding hybrid schemes. The numerical results in Fig.~\ref{Fig1}e show that, over the separation range considered, the effective DM interaction is up to three orders of magnitude stronger than the direct molecular XY exchange underlying recent experimental demonstrations~\cite{entangle_science_bao_23,entangle_science_Holland_23,entangle_nature_iSWAP_25}, enabling entanglement generation on microsecond timescales and fast quantum-information processing with molecular qubits. In the remainder of this work, the molecular pseudospins are treated as qubits, with the basis notation $|0\rangle\equiv|\downarrow\rangle$ and $|1\rangle\equiv|\uparrow\rangle$. We first show how this interaction can be used for fast entanglement generation and universal two-qubit gate operations in the next section.

\subsection*{Fast entangling gates between molecules}\label{Gates}
\begin{figure*}[htbp]
	\centering
	\includegraphics[scale=0.6]{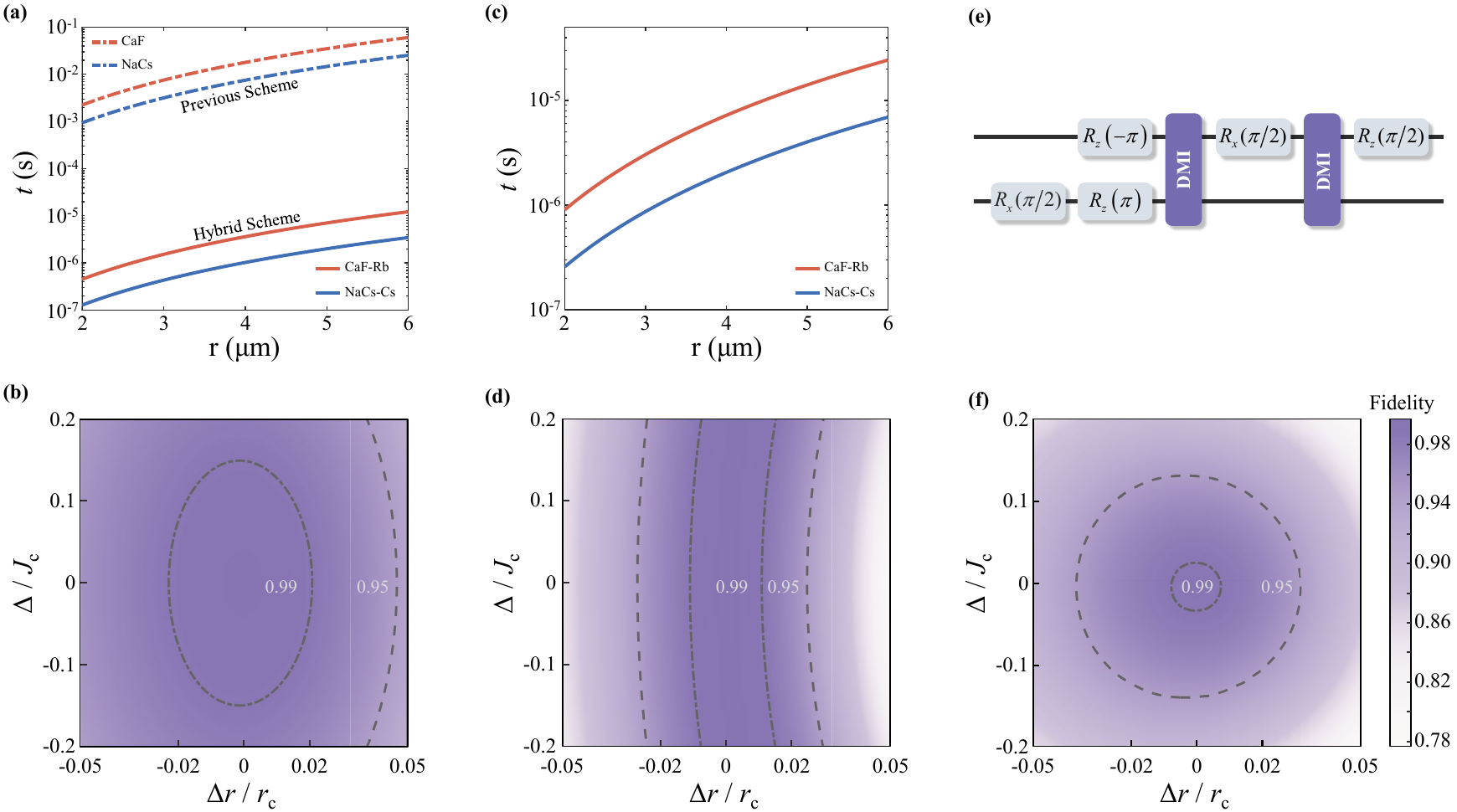}
	\caption{\textbf{Fast entangling gates between molecules.} 
    (a) Bell-state preparation time varying with intermolecular distance $r$ for molecule-only and hybrid schemes.
    (b) Bell-state fidelity as a function of detuning $\Delta$ and atomic-position uncertainty $\Delta r$, obtained from Lindblad master-equation simulations. The chosen parameters are: $r_{1\alpha}= 1.708~\mathrm{\mu m}$, $r_{2\alpha}= 2.292~\mathrm{\mu m}$, corresponding to $J_{1\alpha} = 2\pi\times0.898~\mathrm{MHz}$, $J_{2\alpha} = 2\pi\times0.372~\mathrm{MHz}$.
    (c) SWAP gate time varying with intermolecular distance $r$ and (d) gate fidelity for our hybrid scheme. The calculations use $r_{1\alpha}= 2.292~\mathrm{\mu m}$, $r_{2\alpha}= 1.708~\mathrm{\mu m}$, $r_{1\beta}= 1.708~\mathrm{\mu m}$ and $r_{2\beta}= 2.292~\mathrm{\mu m}$.
    (e) Quantum circuit for constructing a CNOT gate through the DM interaction and single-qubit rotation gates and (f) its performance under realistic imperfections. Unless otherwise specified, simulations use the NaCs--Cs system, with $r_{\mathrm c}=2~\mu\mathrm{m}$ and the corresponding molecule-atom coupling $J_{\mathrm c}=2\pi\times0.559~\mathrm{MHz}$ taken as the reference distance and interaction scale, respectively.}
	\label{Fig2}
\end{figure*}

The strong effective molecular DM spin-exchange interaction enables entangling operations on microsecond timescales. Together with the long coherence times of molecular qubits, such fast operations allow us to achieve large quantum circuit depths for scalable molecular quantum information processing. We demonstrate this capability by first preparing Bell states. Using the single-step evolution $U_1(\alpha)$ (Eq.~\eqref{U1}), an entangling operation can be implemented at $\alpha=\pi/4$, coherently mapping the initial state $\left|01\right\rangle$ or $\left|10\right\rangle$ to the Bell state $\left|\Psi^-\right\rangle = \frac{\sqrt{2}}{2} \left( \left|01\right\rangle - \left|10\right\rangle \right)$ or $-\left|\Psi^+\right\rangle = -\frac{\sqrt{2}}{2} \left( \left|01\right\rangle + \left|10\right\rangle \right)$ after a unitary evolution time $t_1 = 2\pi/\tilde{J}_{\alpha=\pi/4}$. This gives Bell-state preparation times that are orders of magnitude shorter than those in molecule-only schemes based on direct molecular spin-exchange interaction~\cite{entangle_science_bao_23,entangle_science_Holland_23}(Fig.~\ref{Fig2}a). For the NaCs--Cs system at an intermolecular distance $r = r_{1\alpha} + r_{2\alpha} = 4~\mu\mathrm{m}$, the Bell-state preparation time is $ t_1 = 1.029~\mu\mathrm{s}$. 

The gate robustness against detuning fluctuations $\Delta$, atomic-position uncertainty $\Delta r$ and Rydberg decay is evaluated using Lindblad master-equation simulations (see Supplementary Note 2 for details). Here, the atomic-position uncertainty accounts for the finite placement accuracy of the optical tweezers. As shown in Fig.~\ref{Fig2}b, the fidelity is more sensitive to atomic-position uncertainty than to detuning fluctuations. Nevertheless, it remains above 0.95 throughout an extended region of the $(\Delta,\Delta r)$ parameter space, demonstrating  tolerance to both error sources.

By setting $\alpha-\beta=\pi/2$ in Eq.~\eqref{U2}, the initial states $\left|01\right\rangle$ ($\left|10\right\rangle$) are mapped to $\left|10\right\rangle$ ($-\left|01\right\rangle$), realizing a SWAP-like gate through the two-step evolution with a total time $t_1+t_2=2\pi/\tilde{J}_{\alpha}+2\pi/\tilde{J}_{\beta}$. The scaling of the SWAP gate time with intermolecular distance is shown in Fig.~\ref{Fig2}c. For the same NaCs--Cs system with $r=4~\mu\mathrm{m}$, $\alpha=3\pi/4$, and $\beta=\pi/4$, the total gate time is $t=2.058~\mu\mathrm{s}$, corresponding to an effective DM spin-exchange interaction strength $D_{12}^z=2\pi\times0.061~\mathrm{MHz}$. The SWAP gate fidelity is calculated under the same simulation conditions as above. Compared with Bell-state preparation, the two-segment SWAP operation is more sensitive to $\Delta r$ but slightly less sensitive to $\Delta$ (Fig.~\ref{Fig2}d). Consequently, its high-fidelity region is narrower, but the fidelity remains above 0.95 over a substantial range of both error parameters. Furthermore, a standard CNOT gate can be constructed by combining the SWAP gate with several single-qubit rotation gates (Fig.~\ref{Fig2}e). Its fidelity is assessed using the same error model and shows a similar dependence on $\Delta$ and $\Delta r$. Although the high-fidelity region is further reduced, the CNOT gate fidelity remains above 0.95 over a sizeable parameter window (Fig.~\ref{Fig2}f). This construction thus provides a key two-qubit Clifford operation for molecular quantum computation and quantum error correction.

\subsection*{Generation of cluster states via parallel DM interactions}\label{Cluster}
\begin{figure*}[htbp]
	\centering
	\includegraphics[scale=0.55]{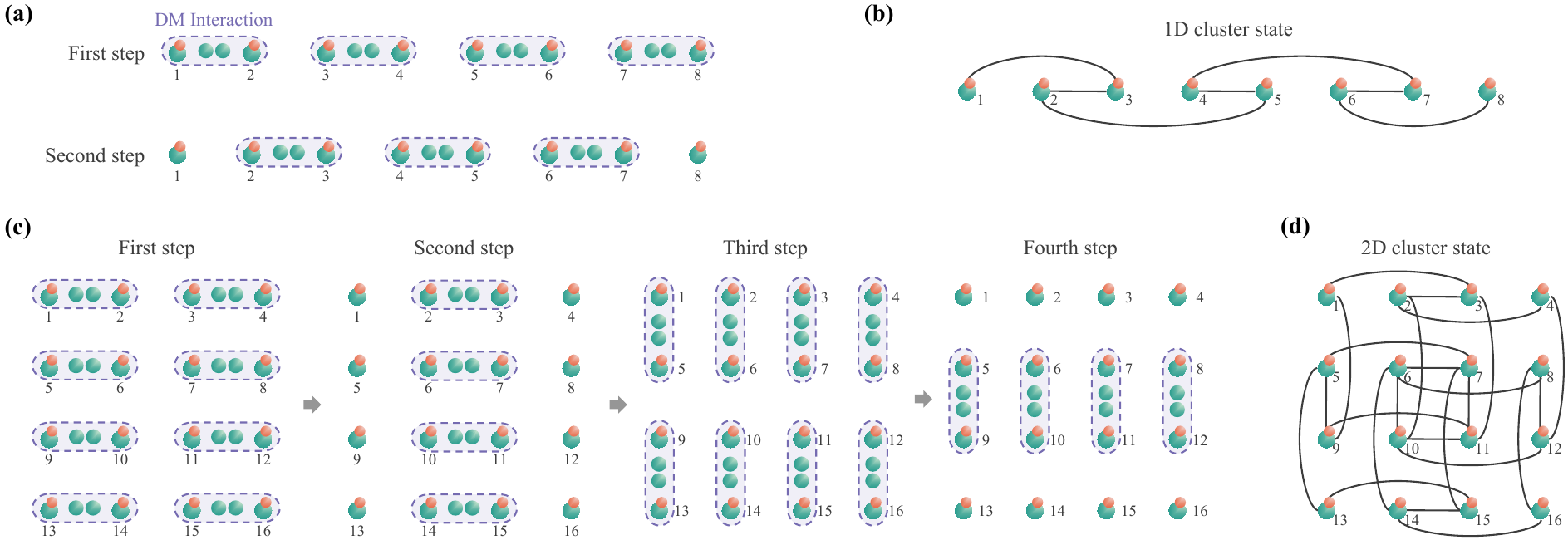}
	\caption{\textbf{Scalable cluster-state generation via parallel DM interactions.}
	(a) Interaction sequence for generating a one-dimensional cluster state. Starting with all molecules initialized in $|+\rangle$, the sequence first activates the DM interaction simultaneously across all odd-indexed bonds, followed by a second parallel round on all even-indexed bonds, thereby fully connecting the nearest-neighbor molecular qubits and producing the linear cluster state shown in (b). (c) Interaction sequence for generating a two-dimensional square-lattice cluster state. Starting with all molecules initialized in $|+\rangle$, the DM interaction is then applied in four layers: two parallel layers on horizontal bonds, followed by two parallel layers on vertical bonds, each arranged into non-overlapping sets to fully establish nearest-neighbor connectivity and produce the two-dimensional cluster state illustrated in (d). The graphs in (b) and (d) are shown in the native qubit ordering generated by the interaction sequence; after reordering the qubits, they are equivalent to a one-dimensional cluster state and a two-dimensional square-lattice cluster state, respectively.}
	\label{Fig3}
\end{figure*}

The fast molecular DM interaction can also be used to generate large-scale multi-particle cluster states, which constitute an important resource for scalable quantum information processing.  In particular, cluster states enable universal measurement-based quantum computation, in which computation is performed through adaptive local measurements on a pre-prepared entangled state, reducing the need for long sequences of coherent gates~\cite{MBQC_one-way_01}. By applying the DM interaction in parallel across non-overlapping molecular pairs, we design two- and four-layer quantum circuits for efficient preparation of one- and two-dimensional cluster states, respectively. 

As an illustrative example, we consider the generation of a three-qubit cluster state. Initializing all molecules in $|+\rangle = (|0\rangle + |1\rangle)/\sqrt{2}$, the DM interaction is sequentially activated on bonds (1-2) and (2-3). Each operation is described by the two-qubit unitary $U^{\mathrm{DM}}=\exp\!\left(-i H_{\textrm{DM}}^z t\right)$ with $D^z_{n,\,n+1}t=\pi/4$, where $t$ is the operation time on each bond:
\begin{equation}
\begin{aligned}
|\phi\rangle_{C_3} &= U^{\textrm{DM}}_{23}U^{\textrm{DM}}_{12}|+\rangle_1|+\rangle_2|+\rangle_3 \\
&= \frac{1}{\sqrt{2}} \left( |+\rangle_1|+\rangle_2|0\rangle_3 + |-\rangle_1|-\rangle_2|1\rangle_3 \right).
\end{aligned}
\end{equation}
This construction readily extends to one-dimensional arrays of arbitrary length by applying parallel DM interactions on non-overlapping bonds. For a general one-dimensional array of $N$ molecules, the cluster state can be prepared using a two-layer interaction sequence illustrated in Fig.~\ref{Fig3}a.
Starting from $|+\rangle^{\otimes N}$, the DM interaction is first activated in parallel on all odd bonds:
\begin{equation}
\prod_{n=1}^{\lfloor N/2 \rfloor}U^{\mathrm{DM}}_{2n-1,\,2n}\; |+\rangle^{\otimes N}\;\longrightarrow\;\bigotimes_{m=1}^{\lfloor N/2 \rfloor}|\phi\rangle_{2m-1,\,2m}. 
\end{equation}
The upper limit $\lfloor N/2 \rfloor$ accounts for both even and odd system sizes: for even $N$, all qubits are paired into $N/2$ disjoint dimers, while for odd $N$ the first layer entangles $(N-1)/2$ bonds, leaving a single unpaired qubit at the end of the chain to be connected in the second layer, which subsequently applies the DM interaction to all even bonds:
\begin{equation}
\prod_{n=1}^{\lfloor (N-1)/2 \rfloor}U^{\mathrm{DM}}_{2n,\,2n+1}\;\bigotimes_{m}|\phi\rangle_{C_{2m-1,\,2m}}
\;=\;|\phi\rangle_{C_N},    
\end{equation}
thereby connecting neighboring dimers into a one-dimensional cluster state, as depicted in Fig.~\ref{Fig3}b. 
The total preparation time is $T_{\mathrm{1D}} = 2t$, independent of system size.

The construction extends straightforwardly to two-dimensional lattices, where the number of required parallel interaction layers is determined by the lattice geometry rather than the system size. For a square lattice, nearest-neighbor connectivity is established through a four-layer interaction sequence, as illustrated in Fig.~\ref{Fig3}c. Starting from all molecules initialized in $|+\rangle$, the DM interaction is applied in parallel on a non-overlapping set of horizontal bonds, followed by a second parallel layer addressing the complementary horizontal bonds. The sequence then applies the DM interaction in parallel on a non-overlapping set of vertical bonds, followed by a final parallel layer on the complementary vertical bonds. These four layers together generate the two-dimensional square-lattice cluster state shown in Fig.~\ref{Fig3}d. Since each layer requires a fixed interaction time $t$, the total preparation time is $T_{\mathrm{2D}} = 4t$. The strong DM interaction brings the molecular cluster-state preparation time into the microsecond regime, while parallel operations on non-overlapping bonds provide a route to fast preparation of large-scale cluster states.

\subsection*{Implementation of non-equilibrium symmetry-protected topological phases}\label{FSPT}
\begin{figure*}[htbp]
	\centering
	\includegraphics[scale=0.525]{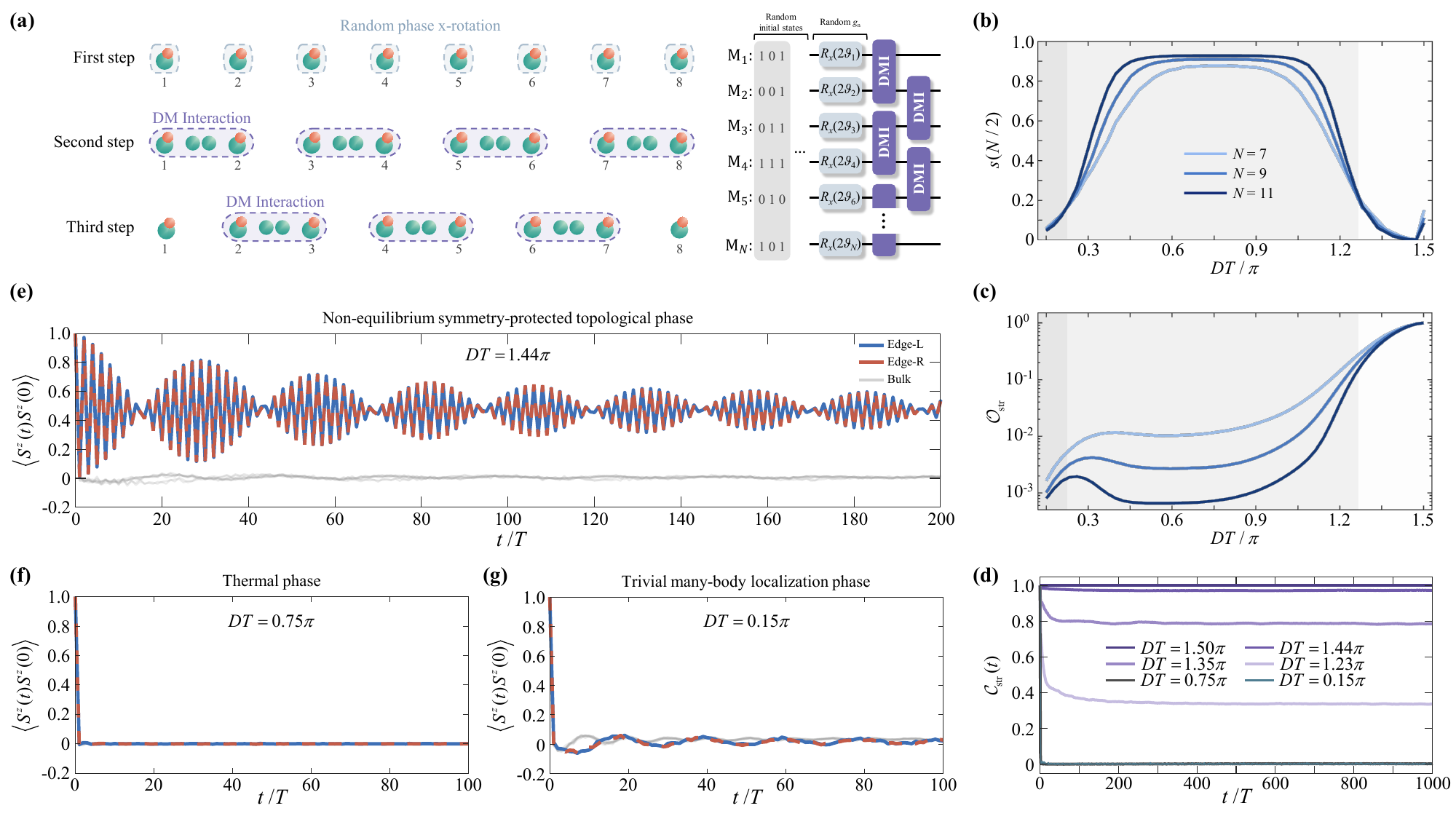}
	\caption{\textbf{Implementation and detection of a non-equilibrium symmetry-protected topological phase via DM interactions.}
    (a) Three-step periodic drive realizing a $\mathbb{Z}_2$ non-equilibrium symmetry-protected topological phase using DM interactions. Disordered onsite rotations promote many-body localization, followed by two staggered DM interaction steps acting on odd and even bonds, respectively; the equivalent quantum circuit representation is shown alongside.
    (b) Normalized bipartite entanglement entropy $s(N/2)$ as a function of interaction parameter $DT$, distinguishing thermal and localized regimes under periodic driving.
    (c) Eigenstate-averaged nonlocal string-order diagnostic $\mathcal{O}_{\mathrm{str}}$, which identifies the non-equilibrium symmetry-protected topological phase within the localized regime.
    (d) Dynamical nonlocal string-order diagnostic $\mathcal{C}_{\mathrm{str}}(t)$, showing persistent nonlocal topological correlations and long-time saturation near the corresponding static value $\mathcal{O}_{\mathrm{str}}$.
    (e--g) Dynamical evolution of the local autocorrelation $\langle S_n^z(t)S_n^z(0)\rangle$ at decreasing values of $DT$, revealing long-lived edge coherence only in the non-equilibrium symmetry-protected topological phase.
    For (d--g), the results are obtained for $N=9$ and averaged over 200 onsite-disorder realizations and 32 randomly sampled computational-basis product states.}
	\label{Fig4}
\end{figure*}
The fast DM interaction permits many driving cycles within the molecular-qubit coherence time, providing access to long-time non-equilibrium many-body dynamics in molecular systems. In particular, non-equilibrium symmetry-protected topological phases have recently attracted considerable interest~\cite{SPT_digital_22,SPT_Sciece_25,SPT_SZM_25,SPT_NRP_26}, owing to their ability to host symmetry-protected long-lived edge coherence, which holds great promise for quantum-information storage and processing~\cite{SPT_QC_prl20,SPT_SZM_25}. Here, we find that the DM interaction can be harnessed to realize a $\mathbb{Z}_2$ non-equilibrium symmetry-protected topological phase under periodic driving. Specifically, we implement a three-step drive (Fig.~\ref{Fig4}a):
\begin{equation}
    H(t)=
    \begin{cases}
    \displaystyle\;\sum_n \;\;g_n S_n^x, & 0\le t<T/3,\\[4pt]
    \displaystyle\sum_{n\in\mathrm{odd}}D\left(S_n^xS_{n+1}^y-S_n^yS_{n+1}^x\right), & T/3\le t<2T/3,\\[4pt]
    \displaystyle\sum_{n\in\mathrm{even}}D\left(S_n^xS_{n+1}^y-S_n^yS_{n+1}^x\right), & 2T/3\le t<T, 
\end{cases}
\end{equation}
where $D\equiv D^z_{n,\,n+1}$. The onsite fields are sampled such that the accumulated rotation angles  $\vartheta_n\equiv g_nT/3\in[-\pi,\pi]$, producing the disordered rotations required to suppress heating through many-body localization. 

The driven phases are first characterized using two eigenstate diagnostics. The normalized bipartite entanglement entropy $s(N/2)$, averaged over the eigenstates $|\alpha\rangle$ of the one-period evolution operator $U(T)$, distinguishes the thermal phase from the localized phases. As shown in Fig.~\ref{Fig4}b, $s$ approaches unity in the thermal regime and scales as $1/N$ in both localized regimes. Because this quantity does not distinguish trivial from topological localization, we next define the dressed string operator~\cite{SPT_Str_92prb} $\mathcal{W}_{[m,n]}=S_m^zU_{[m,n]}S_n^z$, with $U_{[m,n]}$ denoting the one-period evolution restricted to the interval $[m,n]$. The associated eigenstate-averaged string order is $\mathcal{O}_{\mathrm{str}}=2^{-N}\sum_{\alpha}\left|\langle\alpha|\mathcal{W}_{[m,n]}|\alpha\rangle\right|^2$. The resulting $\mathcal O_{\mathrm{str}}$ becomes finite only in the non-equilibrium symmetry-protected topological phase and tends towards unity as $DT$ approaches $1.5\pi$ (Fig.~\ref{Fig4}c).

These eigenstate diagnostics are then related to experimentally accessible dynamics using randomly sampled product states $|\psi_0\rangle=|b_1b_2\cdots b_N\rangle$, with $b_n\in\{0,1\}$. The dynamical string-order parameter $\mathcal{C}_{\mathrm{str}}(m,n;t)=\langle\psi_0|\mathcal{W}_{[m,n]}^\dagger(t)\mathcal{W}_{[m,n]}(0)|\psi_0\rangle$ probes the temporal persistence of nonlocal correlations. In the non-equilibrium symmetry-protected topological regime, $\mathcal C_{\mathrm{str}}(t)$ remains finite and saturates near the corresponding static value $\mathcal O_{\mathrm{str}}$ at long times, whereas it rapidly decays in the thermal and trivial localized regimes (Fig.~\ref{Fig4}d). At the analytically tractable fixed point $DT=1.5\pi$, the one-period evolution factorizes as $U(T)=S_1^zS_N^z\exp\!\left[-i\frac{T}{3}\sum_n g_nS_n^x\right]$, so that the edge operators are fully decoupled from the onsite bulk evolution (see Supplementary Note 5 for a detailed derivation). Close to this point, Fig.~\ref{Fig4}e shows pronounced edge coherence that persists throughout the 200 driving periods at $DT=1.44\pi$, while the bulk correlations rapidly decay. In contrast, no analogous long-lived edge coherence appears in the thermal or trivial localized regimes (Fig.~\ref{Fig4}f,g). 

For the Rydberg-mediated molecular interaction strengths considered here, each driving period is on the order of microseconds, which leads to the observation of hundreds of coherent cycles in the edge-coherence dynamics within the molecular coherence time. In contrast, the direct molecular spin-exchange interaction strength is on the order of tens of hertz at micrometer separations, which greatly limits the number of coherent cycles in the edge-coherence dynamics and thus prevents observation of its long-lived character.

\section*{3.~Discussion and conclusion}
We have introduced a hybrid molecule--atom framework that realizes a Rydberg-atom-mediated spin-exchange interaction Hamiltonian acting solely between molecules, with the auxiliary Rydberg atoms serving as controllable mediators and dynamically decoupling from the molecules at the end of each evolution cycle. By transferring the large transition-dipole scale of Rydberg states to molecular dynamics, this mechanism enhances the interaction strength by up to three orders of magnitude over direct molecular exchange and enables microsecond-scale operations. We use this interaction for rapid Bell-state preparation and two-qubit gates, generation of one- and two-dimensional cluster states, and realization of a non-equilibrium symmetry-protected topological phase with long-lived edge coherence. Combined with the long coherence times and rich internal structure of molecular qubits, these capabilities substantially increase the accessible quantum circuit depth and provide a route toward scalable molecular quantum information processing.

The native DM-$z$ interaction also provides a starting point for spin-chain Hamiltonian engineering. As shown in Supplementary Notes 3 and 4, global and relative single-qubit rotations enable the synthesis of general DM-$xyz$ and XYZ Hamiltonians from the native interaction. Combined with the strong Rydberg-mediated interaction strength, this tunability expands the range of many-body spin-chain Hamiltonians accessible in molecular tweezer arrays, enabling programmable quantum simulation of non-equilibrium dynamics.

\section*{4.~Materials and methods}
\subsection*{Derivation of the effective spin-exchange Hamiltonian}
To derive the effective spin-exchange Hamiltonian (Eq.~(\ref{H})) used in the main text, we first rewrite the microscopic dipole-dipole interaction Hamiltonian (Eq.~(\ref{Hdd})) in the spherical tensor form:
\begin{equation}
	H_{\mathrm{dd}} = -\sqrt{6} \sum_{n=1,2} \frac{\kappa}{r_n^3}
	\sum_{p=-2}^{2} (-1)^p C^2_{-p}(\theta_n,\phi_n) T^2_p(\mathbf{d}_{n},\mathbf{d}_{\mathrm{R}}),
	\label{HddST}
\end{equation}
where $C^2_{-p}(\theta_n,\phi_n)=\sqrt{4\pi/5}\, Y_{2,-p}(\theta_n,\phi_n)$ are reduced spherical harmonics, and $(\theta_n,\phi_n)$ specify the orientation of the molecule-atom separation $\mathbf r_n$ relative to the quantization axis $z$. $T^2_p(\mathbf d_n,\mathbf d_{\mathrm R})$ are rank-2 spherical
tensors coupling the molecular and atomic dipole operators: 
\begin{equation}
    \begin{aligned}
	T_0^2 & = (d_{n}^- d_\mathrm{R}^+ + 2d_{n}^z d_\mathrm{R}^z + d_{n}^+ d_\mathrm{R}^-)/\sqrt{6}, \\
	T_{\pm1}^2 & = (d_{n}^{z} d_\mathrm{R}^\pm + d_{n}^\pm d_\mathrm{R}^z)/\sqrt{2}, \\
	T_{\pm2}^2 & = d_{n}^\pm d_\mathrm{R}^\pm.
    \label{ST}
\end{aligned}
\end{equation}
Here $d_a^z$ and $d_a^\pm$ ($a=n,\mathrm R$) are the spherical components of the corresponding dipole operator, with nonzero matrix elements between internal states satisfying $\Delta m_a=0$ and $\Delta m_a=\pm1$, respectively. Substituting these spherical tensor components into Eq.~(\ref{HddST}), the Hamiltonian can be expressed as
\begin{equation}
\begin{aligned}
H_{\mathrm{dd}}
=&\sum_{n=1,2}\frac{\kappa}{r_n^3}
\left\{\frac{1}{2}\left(1-3\cos^2\theta_n\right)
\left(d_n^-d_{\mathrm R}^+ +2d_n^zd_{\mathrm R}^z+d_n^+d_{\mathrm R}^-\right)\right.\\
&\quad\;+\frac{3}{\sqrt{2}}\sin\theta_n\cos\theta_n\left[e^{-i\phi_n}\left(d_n^zd_{\mathrm R}^+ +d_n^+d_{\mathrm R}^z\right)+\mathrm{H.c.}\right]\\
&\quad\;\left.-\frac{3}{2}\sin^2\theta_n\left[e^{-2i\phi_n}d_n^+d_{\mathrm R}^++\mathrm{H.c.}\right]\right\}.
\end{aligned}
\label{Hdds}
\end{equation}

In the pseudospin encoding, each basis state carries a well-defined total magnetic quantum number $m_{\mathrm{tot}} = m_n + m_{\mathrm{R}}$. Consequently, only the $p=0$ component has non-vanishing matrix elements and yields
\begin{equation}
H_{\mathrm{eff}}^{(p=0)}
=\sum_{n=1,2}V_{\mathrm{dd}}(r_n)\left[J_zS_n^zS_{\mathrm R}^z+\frac{J_\perp}{2}
\left(S_n^+S_{\mathrm R}^-+S_n^-S_{\mathrm R}^+\right)
\right],
\end{equation}
where
$V_{\mathrm{dd}}(r_n)=\frac{\kappa}{r_n^3}(1-3\cos^2\theta_n)$ is a geometry-dependent factor. In general, projection onto the pseudospin manifold yields this XXZ-type interaction containing both exchange and Ising contributions. The exchange coupling $J_{\perp}$ is determined by the transition dipole moments, whereas the Ising coupling $J_z$ originates from the permanent dipole moments of the corresponding pseudospin states:
$$J_{\perp}\!\propto\! d^{\uparrow\downarrow}_{n} d^{\uparrow\downarrow}_{\mathrm{R}},\,\,
    J_{z}\!\propto d^{\uparrow\uparrow}_{n} d^{\uparrow\uparrow}_{\mathrm{R}}
    + d^{\downarrow\downarrow}_{n} d^{\downarrow\downarrow}_{\mathrm{R}}
    - d^{\uparrow\uparrow}_{n} d^{\downarrow\downarrow}_{\mathrm{R}}
    - d^{\downarrow\downarrow}_{n} d^{\uparrow\uparrow}_{\mathrm{R}}.$$
Therefore, the Ising interaction disappears whenever the two pseudospin states possess identical permanent dipole moments. In the implementation considered here, the rotational and Rydberg pseudospin states are chosen such that (i) the two states within each doublet have identical permanent dipole moments and (ii) the molecular and Rydberg exchange transitions are mutually resonant. Consequently, the first condition gives $J_z=0$, while the second ensures that the remaining exchange interaction is resonant. The effective Hamiltonian therefore reduces to the pure spin-exchange form in Eq.~(\ref{H}). The corresponding distance-dependent exchange amplitudes are
\begin{equation}
	J_{n} = \frac{\kappa  d^{\uparrow\downarrow}_{n}  d^{\uparrow\downarrow}_{\mathrm{R}}}{r^{3}_{n}} \left(1-3 \cos ^{2} \theta_n\right).
	\label{Jgen}
\end{equation}
For the geometry considered in the main text, the quantization axis $z$ is perpendicular to the molecular $x$--$y$ plane (Fig.~\ref{Fig1}b, inset), so that $\theta_n=\pi/2$ and Eq.~(\ref{Jgen}) reduces to Eq.~(\ref{J}).

\section*{Conflict of interest}
The authors declare that they have no conflict of interest.

\section*{Acknowledgments}
This work is supported by the National Key Research and Development Program of China (Grant No. 2022YFA1404201), National Natural Science Foundation of China (NSFC) (Grant No. 62325505, No. 12474361, No. 12074234), Changjiang Scholars and Innovative Research Team in University of Ministry of Education of China (PCSIRT)($IRT\_17R70$), Fund for Shanxi 1331 Project Key Subjects Construction, 111 Project (D18001) and Fundamental Research Program of Shanxi Province (Grant No. 202303021223005). Weibin Li acknowledges support from The Engineering and Physical Sciences Research Council (EPSRC) through Grant No. EP/W015641/1. Chi Zhang acknowledges support from EPSRC through Award Number UKRI3888.

\section*{Author contributions}
These authors contributed equally: Yunqing Jiao, Jin-Zhu Jiang.

Chi Zhang, Weibin Li, and Feng Mei conceived the project. Yunqing Jiao, Jin-Zhu Jiang, and Bo-Wen Guan performed the theoretical calculations. Yunqing Jiao, Jin-Zhu Jiang, Chi Zhang, Weibin Li, and Feng Mei carried out the theoretical analysis. Jie Ma and Liantuan
Xiao provided valuable discussions. Suotang Jia supervised the project. All authors discussed the results, contributed to the data analysis, and co-wrote the paper.

\bibliographystyle{sb-vancouver3}
\bibliography{refv4}

@article{NPreview_Chemistry_24,
	title = {Ultracold chemistry as a testbed for few-body physics},
	volume = {20},
	issn = {1745-2473, 1745-2481},
	doi = {10.1038/s41567-024-02467-3},
	language = {en},
	number = {5},
	urldate = {2026-06-09},
	journal = {Nat Phys},
	author = {Karman, Tijs and Tomza, Michał and Pérez-Ríos, Jesús},
	month = may,
	year = {2024},
	pages = {722--729},
}

@article{NPreview_Maniandcooling_24,
	title = {Quantum state manipulation and cooling of ultracold molecules},
	volume = {20},
	issn = {1745-2473, 1745-2481},
	doi = {10.1038/s41567-024-02423-1},
	language = {en},
	number = {5},
	urldate = {2026-06-09},
	journal = {Nat Phys},
	author = {Langen, Tim and Valtolina, Giacomo and Wang, Dajun and Ye, Jun},
	month = may,
	year = {2024},
	pages = {702--712},
}

@article{NPreview_QCQS_24,
	title = {Quantum computation and quantum simulation with ultracold molecules},
	volume = {20},
	issn = {1745-2473, 1745-2481},
	doi = {10.1038/s41567-024-02453-9},
	language = {en},
	number = {5},
	urldate = {2026-06-09},
	journal = {Nat Phys},
	author = {Cornish, Simon L. and Tarbutt, Michael R. and Hazzard, Kaden R. A.},
	month = may,
	year = {2024},
	pages = {730--740},
}

@article{NPreview_Metrology_24,
	title = {Quantum sensing and metrology for fundamental physics with molecules},
	volume = {20},
	issn = {1745-2473, 1745-2481},
	doi = {10.1038/s41567-024-02499-9},
	language = {en},
	number = {5},
	urldate = {2026-06-09},
	journal = {Nat Phys},
	author = {DeMille, David and Hutzler, Nicholas R. and Rey, Ana Maria and Zelevinsky, Tanya},
	month = may,
	year = {2024},
	pages = {741--749},
}

@article{tweezer_19,
	title = {An optical tweezer array of ultracold molecules},
	volume = {365},
	issn = {0036-8075, 1095-9203},
	doi = {10.1126/science.aax1265},
	language = {en},
	number = {6458},
	urldate = {2026-06-09},
	journal = {Science},
	author = {Anderegg, Loïc and Cheuk, Lawrence W. and Bao, Yicheng and Burchesky, Sean and Ketterle, Wolfgang and Ni, Kang-Kuen and Doyle, John M.},
	month = sep,
	year = {2019},
	pages = {1156--1158},
}

@article{tweezer_20,
	title = {Forming a {Single} {Molecule} by {Magnetoassociation} in an {Optical} {Tweezer}},
	volume = {124},
	issn = {0031-9007, 1079-7114},
	doi = {10.1103/PhysRevLett.124.253401},
	language = {en},
	number = {25},
	urldate = {2026-06-09},
	journal = {Phys Rev Lett},
	author = {Zhang, Jessie T. and Yu, Yichao and Cairncross, William B. and Wang, Kenneth and Picard, Lewis R. B. and Hood, Jonathan D. and Lin, Yen-Wei and Hutson, Jeremy M. and Ni, Kang-Kuen},
	month = jun,
	year = {2020},
	pages = {253401},
}

@article{tweezer_ployatomic_24,
	title = {An optical tweezer array of ultracold polyatomic molecules},
	volume = {628},
	issn = {0028-0836, 1476-4687},
	doi = {10.1038/s41586-024-07199-1},
	language = {en},
	number = {8007},
	urldate = {2026-06-09},
	journal = {Nature},
	author = {Vilas, Nathaniel B. and Robichaud, Paige and Hallas, Christian and Li, Grace K. and Anderegg, Loïc and Doyle, John M.},
	month = apr,
	year = {2024},
	pages = {282--286},
}

@article{tweezer_enhanced_24,
	title = {Enhanced {Quantum} {Control} of {Individual} {Ultracold} {Molecules} {Using} {Optical} {Tweezer} {Arrays}},
	volume = {5},
	issn = {2691-3399},
	doi = {10.1103/PRXQuantum.5.020333},
	language = {en},
	number = {2},
	urldate = {2026-06-09},
	journal = {PRX Quantum},
	author = {Ruttley, Daniel K. and Guttridge, Alexander and Hepworth, Tom R. and Cornish, Simon L.},
	month = may,
	year = {2024},
	pages = {020333},
}

@article{tweezer_site-selective_24,
	title = {Site-{Selective} {Preparation} and {Multistate} {Readout} of {Molecules} in {Optical} {Tweezers}},
	volume = {5},
	issn = {2691-3399},
	doi = {10.1103/PRXQuantum.5.020344},
	language = {en},
	number = {2},
	urldate = {2026-06-09},
	journal = {PRX Quantum},
	author = {Picard, Lewis R. B. and Patenotte, Gabriel E. and Park, Annie J. and Gebretsadkan, Samuel F. and Ni, Kang-Kuen},
	month = may,
	year = {2024},
	pages = {020344},
}

@article{tweezer_JingZ_14,
	title = {Production of {Feshbach} molecules induced by spin–orbit coupling in {Fermi} gases},
	volume = {10},
	issn = {1745-2473, 1745-2481},
	doi = {10.1038/nphys2824},
	language = {en},
	number = {2},
	urldate = {2026-07-15},
	journal = {Nat Phys},
	author = {Fu, Zhengkun and Huang, Lianghui and Meng, Zengming and Wang, Pengjun and Zhang, Long and Zhang, Shizhong and Zhai, Hui and Zhang, Peng and Zhang, Jing},
	month = feb,
	year = {2014},
	pages = {110--115},
}

@article{tweezer_storage_21,
	title = {Robust storage qubits in ultracold polar molecules},
	volume = {17},
	issn = {1745-2473, 1745-2481},
	doi = {10.1038/s41567-021-01328-7},
	language = {en},
	number = {10},
	urldate = {2026-06-09},
	journal = {Nat Phys},
	author = {Gregory, Philip D. and Blackmore, Jacob A. and Bromley, Sarah L. and Hutson, Jeremy M. and Cornish, Simon L.},
	month = oct,
	year = {2021},
	pages = {1149--1153},
}

@article{tweezer_preparation_22,
	title = {Preparation of $^{\textrm{87}}${Rb} and $^{\textrm{133}}${Cs} in the motional ground state of a single optical tweezer},
	volume = {24},
	issn = {1367-2630},
	doi = {10.1088/1367-2630/ac95b9},
	language = {en},
	number = {10},
	urldate = {2026-06-09},
	journal = {New J Phys},
	author = {Spence, S and Brooks, R V and Ruttley, D K and Guttridge, A and Cornish, Simon L},
	month = oct,
	year = {2022},
	pages = {103022},
}

@article{tweezer_Zhan_20,
	title = {Coherently forming a single molecule in an optical trap},
	volume = {370},
	issn = {0036-8075, 1095-9203},
	doi = {10.1126/science.aba7468},
	language = {en},
	number = {6514},
	urldate = {2026-06-11},
	journal = {Science},
	author = {He, Xiaodong and Wang, Kunpeng and Zhuang, Jun and Xu, Peng and Gao, Xiang and Guo, Ruijun and Sheng, Cheng and Liu, Min and Wang, Jin and Li, Jiaming and Shlyapnikov, G. V. and Zhan, Mingsheng},
	month = oct,
	year = {2020},
	pages = {331--335},
}

@article{tweezer_spin1_25,
	title = {Long-lived multilevel coherences and spin-1 dynamics encoded in the rotational states of ultracold molecules},
	volume = {16},
	issn = {2041-1723},
	doi = {10.1038/s41467-025-62275-y},
	language = {en},
	number = {1},
	urldate = {2026-06-11},
	journal = {Nat Commun},
	author = {Hepworth, Tom R. and Ruttley, Daniel K. and Von Gierke, Fritz and Gregory, Philip D. and Guttridge, Alexander and Cornish, Simon L.},
	month = aug,
	year = {2025},
	pages = {7131},
}

@article{tweezer_raman_24,
	title = {Raman sideband cooling of molecules in an optical tweezer array},
	volume = {20},
	issn = {1745-2473, 1745-2481},
	doi = {10.1038/s41567-023-02346-3},
	language = {en},
	number = {3},
	urldate = {2026-06-11},
	journal = {Nat Phys},
	author = {Lu, Yukai and Li, Samuel J. and Holland, Connor M. and Cheuk, Lawrence W.},
	month = mar,
	year = {2024},
	pages = {389--394},
}

@article{tweezer_raman3D_24,
	title = {Raman {Sideband} {Cooling} of {Molecules} in an {Optical} {Tweezer} {Array} to the {3D} {Motional} {Ground} {State}},
	volume = {14},
	issn = {2160-3308},
	doi = {10.1103/PhysRevX.14.031002},
	language = {en},
	number = {3},
	urldate = {2026-06-11},
	journal = {Phys Rev X},
	author = {Bao, Yicheng and Yu, Scarlett S. and You, Jiaqi and Anderegg, Loïc and Chae, Eunmi and Ketterle, Wolfgang and Ni, Kang-Kuen and Doyle, John M.},
	month = jul,
	year = {2024},
	pages = {031002},
}

@article{tweezer_3DMOT_BYprl24,
	title = {Three-{Dimensional} {Magneto}-{Optical} {Trapping} of {Barium} {Monofluoride}},
	volume = {133},
	issn = {0031-9007, 1079-7114},
	doi = {10.1103/PhysRevLett.133.143404},
	language = {en},
	number = {14},
	urldate = {2026-08-14},
	journal = {Phys Rev Lett},
	author = {Zeng, Zixuan and Deng, Shuhua and Yang, Shoukang and Yan, Bo},
	month = oct,
	year = {2024},
	pages = {143404},
}

@article{tweezer_MBSD_Cheuk26,
	title = {Probing {Coherent} {Many}-{Body} {Spin} {Dynamics} in a {Molecular} {Tweezer} {Array} {Quantum} {Simulator}},
	doi = {10.48550/arXiv.2603.19090},
	language = {en},
	urldate = {2026-08-14},
	journal = {arXiv:2603.19090},
	author = {Lu, Yukai and Holland, Connor M. and Welsh, Callum L. and Chen, Xing-Yan and Cheuk, Lawrence W.},
	month = mar,
	year = {2026}
}

@article{Shield_TaoS_26NP,
	title = {Bose–{Einstein} condensate of ultracold sodium–rubidium molecules with tunable dipolar interactions},
	issn = {1745-2473, 1745-2481},
	doi = {10.1038/s41567-026-03362-9},
	language = {en},
	urldate = {2026-07-10},
	journal = {Nat Phys},
	author = {Shi, Zhaopeng and Huang, Zerong and Deng, Fulin and Jin, Wei-Jian and Yi, Su and Shi, Tao and Wang, Dajun},
	month = jul,
	year = {2026},
}

@article{Shield_XCui_26L,
	title = {Universal {Bound} {States} with {Bose}-{Fermi} {Duality} in {Microwave}-{Shielded} {Ultracold} {Molecules}},
	volume = {136},
	issn = {0031-9007, 1079-7114},
	doi = {10.1103/hcwf-tk6c},
	language = {en},
	number = {4},
	urldate = {2026-07-08},
	journal = {Phys Rev Lett},
	author = {Shi, Tingting and Wang, Haitian and Cui, Xiaoling},
	month = jan,
	year = {2026},
	pages = {043402},
}

@article{Shield_TaoS_23L,
	title = {Effective {Potential} and {Superfluidity} of {Microwave}-{Shielded} {Polar} {Molecules}},
	volume = {130},
	issn = {0031-9007, 1079-7114},
	doi = {10.1103/PhysRevLett.130.183001},
	language = {en},
	number = {18},
	urldate = {2026-07-08},
	journal = {Phys Rev Lett},
	author = {Deng, Fulin and Chen, Xing-Yan and Luo, Xin-Yu and Zhang, Wenxian and Yi, Su and Shi, Tao},
	month = may,
	year = {2023},
	pages = {183001},
}

@article{Shield_TaoS_25L,
	title = {Bose-{Einstein} {Condensates} of {Microwave}-{Shielded} {Polar} {Molecules}},
	volume = {134},
	issn = {0031-9007, 1079-7114},
	doi = {10.1103/b8y9-yvz9},
	language = {en},
	number = {23},
	urldate = {2026-07-08},
	journal = {Phys Rev Lett},
	author = {Jin, Wei-Jian and Deng, Fulin and Yi, Su and Shi, Tao},
	month = jun,
	year = {2025},
	pages = {233003},
}

@article{Shield_IBloch_22Nat,
	title = {Evaporation of microwave-shielded polar molecules to quantum degeneracy},
	volume = {607},
	issn = {0028-0836, 1476-4687},
	doi = {10.1038/s41586-022-04900-0},
	language = {en},
	number = {7920},
	urldate = {2026-07-08},
	journal = {Nature},
	author = {Schindewolf, Andreas and Bause, Roman and Chen, Xing-Yan and Duda, Marcel and Karman, Tijs and Bloch, Immanuel and Luo, Xin-Yu},
	month = jul,
	year = {2022},
	pages = {677--681},
}

@article{Shield_Doyle_21Sci,
	title = {Observation of microwave shielding of ultracold molecules},
	volume = {373},
	issn = {0036-8075, 1095-9203},
	doi = {10.1126/science.abg9502},
	language = {en},
	number = {6556},
	urldate = {2026-07-08},
	journal = {Science},
	author = {Anderegg, Loïc and Burchesky, Sean and Bao, Yicheng and Yu, Scarlett S. and Karman, Tijs and Chae, Eunmi and Ni, Kang-Kuen and Ketterle, Wolfgang and Doyle, John M.},
	month = aug,
	year = {2021},
	pages = {779--782},
}

@article{Shield_Scatter_prl18,
  title = {Controlling the Scattering Length of Ultracold Dipolar Molecules},
  author = {Lassabli\`ere, Lucas and Qu\'em\'ener, Goulven},
  journal = {Phys Rev Lett},
  volume = {121},
  issue = {16},
  pages = {163402},
  numpages = {6},
  year = {2018},
  month = {Oct},
  publisher = {American Physical Society},
  doi = {10.1103/PhysRevLett.121.163402},
}

@article{Shield_MSPM_prl18,
  title = {Microwave Shielding of Ultracold Polar Molecules},
  author = {Karman, Tijs and Hutson, Jeremy M.},
  journal = {Phys Rev Lett},
  volume = {121},
  issue = {16},
  pages = {163401},
  numpages = {5},
  year = {2018},
  month = {Oct},
  publisher = {American Physical Society},
  doi = {10.1103/PhysRevLett.121.163401},
}

@article{Shield_field-linked_N23,
	title = {Field-linked resonances of polar molecules},
	volume = {614},
	issn = {0028-0836, 1476-4687},
	doi = {10.1038/s41586-022-05651-8},
	language = {en},
	number = {7946},
	urldate = {2026-08-11},
	journal = {Nature},
	author = {Chen, Xing-Yan and Schindewolf, Andreas and Eppelt, Sebastian and Bause, Roman and Duda, Marcel and Biswas, Shrestha and Karman, Tijs and Hilker, Timon and Bloch, Immanuel and Luo, Xin-Yu},
	month = feb,
	year = {2023},
	pages = {59--63},
}

@article{Shield_tetratomic_N24,
	title = {Ultracold field-linked tetratomic molecules},
	volume = {626},
	issn = {0028-0836, 1476-4687},
	doi = {10.1038/s41586-023-06986-6},
	language = {en},
	number = {7998},
	urldate = {2026-08-11},
	journal = {Nature},
	author = {Chen, Xing-Yan and Biswas, Shrestha and Eppelt, Sebastian and Schindewolf, Andreas and Deng, Fulin and Shi, Tao and Yi, Su and Hilker, Timon A. and Bloch, Immanuel and Luo, Xin-Yu},
	month = feb,
	year = {2024},
	pages = {283--287},
}

@article{entangle_PRL02_DeMille,
  title = {Quantum Computation with Trapped Polar Molecules},
  author = {DeMille, D.},
  journal = {Phys Rev Lett},
  volume = {88},
  issue = {6},
  pages = {067901},
  numpages = {4},
  year = {2002},
  month = {Jan},
  publisher = {American Physical Society},
  doi = {10.1103/PhysRevLett.88.067901},
}

@article{entangle_science_bao_23,
	title = {Dipolar spin-exchange and entanglement between molecules in an optical tweezer array},
	volume = {382},
	issn = {0036-8075, 1095-9203},
	doi = {10.1126/science.adf8999},
	language = {en},
	number = {6675},
	urldate = {2026-06-09},
	journal = {Science},
	author = {Bao, Yicheng and Yu, Scarlett S. and Anderegg, Loïc and Chae, Eunmi and Ketterle, Wolfgang and Ni, Kang-Kuen and Doyle, John M.},
	month = dec,
	year = {2023},
	pages = {1138--1143},
}

@article{entangle_science_Holland_23,
	title = {On-demand entanglement of molecules in a reconfigurable optical tweezer array},
	volume = {382},
	issn = {0036-8075, 1095-9203},
	doi = {10.1126/science.adf4272},
	language = {en},
	number = {6675},
	urldate = {2026-06-09},
	journal = {Science},
	author = {Holland, Connor M. and Lu, Yukai and Cheuk, Lawrence W.},
	month = dec,
	year = {2023},
	pages = {1143--1147},
}

@article{entangle_KKNi_CS18,
	title = {Dipolar exchange quantum logic gate with polar molecules},
	volume = {9},
	issn = {2041-6520, 2041-6539},
	doi = {10.1039/C8SC02355G},
	language = {en},
	number = {33},
	urldate = {2026-07-15},
	journal = {Chem Sci},
	author = {Ni, Kang-Kuen and Rosenband, Till and Grimes, David D.},
	year = {2018},
	pages = {6830--6838},
}

@article{entangle_nature_iSWAP_25,
	title = {Entanglement and {iSWAP} gate between molecular qubits},
	volume = {637},
	issn = {0028-0836, 1476-4687},
	doi = {10.1038/s41586-024-08177-3},
	language = {en},
	number = {8047},
	urldate = {2026-06-09},
	journal = {Nature},
	author = {Picard, Lewis R. B. and Park, Annie J. and Patenotte, Gabriel E. and Gebretsadkan, Samuel and Wellnitz, David and Rey, Ana Maria and Ni, Kang-Kuen},
	month = jan,
	year = {2025},
	pages = {821--826},
}

@article{entangle_nature_longlived_25,
	title = {Long-lived entanglement of molecules in magic-wavelength optical tweezers},
	volume = {637},
	issn = {0028-0836, 1476-4687},
	doi = {10.1038/s41586-024-08365-1},
	language = {en},
	number = {8047},
	urldate = {2026-06-09},
	journal = {Nature},
	author = {Ruttley, Daniel K. and Hepworth, Tom R. and Guttridge, Alexander and Cornish, Simon L.},
	month = jan,
	year = {2025},
	pages = {827--832},
}

@article{entangle_squeezing_26,
	title = {Creating and {Probing} {Spin}-{Squeezed} {States} of {Molecules}},
	doi = {10.48550/arXiv.2606.02500},
	language = {en},
	urldate = {2026-06-11},
	journal = {arXiv:2606.02500},
	author = {Holland, Connor M. and Welsh, Callum L. and Lu, Yukai and Wellnitz, David and Chen, Xing-Yan and Rey, Ana Maria and Cheuk, Lawrence W.},
	year = {2026},
}

@article{Rydberg_review_21,
	title = {Quantum science with optical tweezer arrays of ultracold atoms and molecules},
	volume = {17},
	issn = {1745-2473, 1745-2481},
	doi = {10.1038/s41567-021-01357-2},
	language = {en},
	number = {12},
	urldate = {2026-06-15},
	journal = {Nat Phys},
	author = {Kaufman, Adam M. and Ni, Kang-Kuen},
	month = dec,
	year = {2021},
	pages = {1324--1333},
}

@article{Rydberg_review_20,
	title = {Many-body physics with individually controlled {Rydberg} atoms},
	volume = {16},
	issn = {1745-2473, 1745-2481},
	doi = {10.1038/s41567-019-0733-z},
	language = {en},
	number = {2},
	urldate = {2026-06-15},
	journal = {Nat Phys},
	author = {Browaeys, Antoine and Lahaye, Thierry},
	month = feb,
	year = {2020},
	pages = {132--142},
}

@article{hybrid_readout_Yelinpra16,
  title = {Rydberg-atom-mediated nondestructive readout of collective rotational states in polar-molecule arrays},
  author = {Kuznetsova, Elena and Rittenhouse, Seth T. and Sadeghpour, H. R. and Yelin, Susanne F.},
  journal = {Phys Rev A},
  volume = {94},
  issue = {3},
  pages = {032325},
  numpages = {21},
  year = {2016},
  month = {Sep},
  publisher = {American Physical Society},
  doi = {10.1103/PhysRevA.94.032325},
}

@article{hybrid_blockade_23,
	title = {Observation of {Rydberg} {Blockade} {Due} to the {Charge}-{Dipole} {Interaction} between an {Atom} and a {Polar} {Molecule}},
	volume = {131},
	issn = {0031-9007, 1079-7114},
	doi = {10.1103/PhysRevLett.131.013401},
	language = {en},
	number = {1},
	urldate = {2026-06-09},
	journal = {Phys Rev Lett},
	author = {Guttridge, Alexander and Ruttley, Daniel K. and Baldock, Archie C. and González-Férez, Rosario and Sadeghpour, H. R. and Adams, C. S. and Cornish, Simon L.},
	month = jul,
	year = {2023},
	pages = {013401},
}

@article{hybrid_probing_25,
	title = {Probing {Dipolar} {Interactions} between {Rydberg} {Atoms} and {Ultracold} {Polar} {Molecules}},
	volume = {135},
	issn = {0031-9007, 1079-7114},
	doi = {10.1103/48rk-sxfs},
	language = {en},
	number = {15},
	urldate = {2026-06-09},
	journal = {Phys Rev Lett},
	author = {Zhu, Lingbang and Luke, Jeshurun and Shaham, Roy and Liu, Yi-Xiang and Ni, Kang-Kuen},
	month = oct,
	year = {2025},
	pages = {153001},
}

@article{hybrid_PRXquantum_enriching_22,
	title = {Enriching the {Quantum} {Toolbox} of {Ultracold} {Molecules} with {Rydberg} {Atoms}},
	volume = {3},
	issn = {2691-3399},
	doi = {10.1103/PRXQuantum.3.030339},
	language = {en},
	number = {3},
	urldate = {2026-06-09},
	journal = {PRX Quantum},
	author = {Wang, Kenneth and Williams, Conner P. and Picard, Lewis R.B. and Yao, Norman Y. and Ni, Kang-Kuen},
	month = sep,
	year = {2022},
	pages = {030339},
}

@article{hybrid_PRXquantum_ChiZhang_22,
	title = {Quantum {Computation} in a {Hybrid} {Array} of {Molecules} and {Rydberg} {Atoms}},
	volume = {3},
	issn = {2691-3399},
	doi = {10.1103/PRXQuantum.3.030340},
	language = {en},
	number = {3},
	urldate = {2026-06-09},
	journal = {PRX Quantum},
	author = {Zhang, Chi and Tarbutt, M.R.},
	month = sep,
	year = {2022},
	pages = {030340},
}

@article{hybrid_Harness_26,
	title = {Harnessing resonant dipolar interactions in a hybrid atom-molecule quantum system},
	doi = {10.48550/arXiv.2607.15976},
	language = {en},
	urldate = {2026-07-21},
	journal = {arXiv:2607.15976},
	author = {Ruttley, Daniel K. and Hepworth, Tom R. and García-Garrido, Juan M. and Rich, Caleb J. H. and González-Férez, Rosario and Guttridge, Alexander and Cornish, Simon L.},
	year = {2026},
}

@article{hybrid_readout_26,
	title = {Simultaneous nondestructive measurement of many polar molecules using {Rydberg} atoms},
	doi = {10.48550/arXiv.2601.08921},
	language = {en},
	urldate = {2026-06-09},
	journal = {arXiv:2601.08921},
	author = {Young, Jeremy T. and Ni, Kang-Kuen and Gorshkov, Alexey V.},
	year = {2026},
}

@article{hybrid_GHZ_26,
	title = {Quantum logic control and entanglement in hybrid atom-molecule arrays},
	doi = {10.48550/arXiv.2602.12909},
	language = {en},
	urldate = {2026-06-10},
	journal = {arXiv:2602.12909},
	author = {Zhang, Chi and Murciano, Sara and Tantivasadakarn, Nathanan and Finkelstein, Ran},
	month = feb,
	year = {2026},
}

@article{hybrid_CNOT_26,
  title = {Multipartite Controlled-{NOT} Gates Using Molecules and {Rydberg} Atoms},
  author = {Bai, Yi-Han and Wei, Yue and Zhang, Chi and Li, Weibin and Shao, Xiao-Qiang},
  journal = {Chin Phys Lett},
  volume = {43},
  pages = {080302},
  year = {2026},
  doi = {10.1088/0256-307X/43/8/080302},
}

@article{MBQC_one-way_01,
	title = {A {One}-{Way} {Quantum} {Computer}},
	volume = {86},
	issn = {0031-9007, 1079-7114},
	doi = {10.1103/PhysRevLett.86.5188},
	language = {en},
	number = {22},
	urldate = {2026-06-09},
	journal = {Phys Rev Lett},
	author = {Raussendorf, Robert and Briegel, Hans J.},
	month = may,
	year = {2001},
	pages = {5188--5191},
}

@article{DDI_gorshkovPRL_11,
	title = {Tunable {Superfluidity} and {Quantum} {Magnetism} with {Ultracold} {Polar} {Molecules}},
	volume = {107},
	issn = {0031-9007, 1079-7114},
	doi = {10.1103/PhysRevLett.107.115301},
	language = {en},
	number = {11},
	urldate = {2026-06-17},
	journal = {Phys Rev Lett},
	author = {Gorshkov, Alexey V. and Manmana, Salvatore R. and Chen, Gang and Ye, Jun and Demler, Eugene and Lukin, Mikhail D. and Rey, Ana Maria},
	month = sep,
	year = {2011},
	pages = {115301},
}

@article{DDI_gorshkovPRA_11,
	title = {Quantum magnetism with polar alkali-metal dimers},
	volume = {84},
	issn = {1050-2947, 1094-1622},
	doi = {10.1103/PhysRevA.84.033619},
	language = {en},
	number = {3},
	urldate = {2026-06-17},
	journal = {Phys Rev A},
	author = {Gorshkov, Alexey V. and Manmana, Salvatore R. and Chen, Gang and Demler, Eugene and Lukin, Mikhail D. and Rey, Ana Maria},
	month = sep,
	year = {2011},
	pages = {033619},
}

@article{DDI_BoYanNature_13,
	title = {Observation of dipolar spin-exchange interactions with lattice-confined polar molecules},
	volume = {501},
	issn = {0028-0836, 1476-4687},
	doi = {10.1038/nature12483},
	language = {en},
	number = {7468},
	urldate = {2026-06-17},
	journal = {Nature},
	author = {Yan, Bo and Moses, Steven A. and Gadway, Bryce and Covey, Jacob P. and Hazzard, Kaden R. A. and Rey, Ana Maria and Jin, Deborah S. and Ye, Jun},
	month = sep,
	year = {2013},
	pages = {521--525},
}

@article{DDI_JYeNat_24,
	title = {Two-axis twisting using {Floquet}-engineered {XYZ} spin models with polar molecules},
	volume = {633},
	issn = {0028-0836, 1476-4687},
	doi = {10.1038/s41586-024-07883-2},
	language = {en},
	number = {8029},
	urldate = {2026-07-16},
	journal = {Nature},
	author = {Miller, Calder and Carroll, Annette N. and Lin, Junyu and Hirzler, Henrik and Gao, Haoyang and Zhou, Hengyun and Lukin, Mikhail D. and Ye, Jun},
	month = sep,
	year = {2024},
	pages = {332--337},
}

@article{DDI_JYeSci_25,
    author = {Annette N. Carroll  and Henrik Hirzler  and Calder Miller  and David Wellnitz  and Sean R. Muleady  and Junyu Lin  and Krzysztof P. Zamarski  and Reuben R. W. Wang  and John L. Bohn  and Ana Maria Rey  and Jun Ye },
    title = {Observation of generalized t-J spin dynamics with tunable dipolar interactions},
    journal = {Science},
    volume = {388},
    number = {6745},
    pages = {381-386},
    year = {2025},
    doi = {10.1126/science.adq0911},
}

@article{SPT_Str_92prb,
  title = {Hidden ${\mathrm{Z}}_{2}$\ifmmode\times\else\texttimes\fi{}${\mathrm{Z}}_{2}$ symmetry breaking in Haldane-gap antiferromagnets},
  author = {Kennedy, Tom and Tasaki, Hal},
  journal = {Phys Rev B},
  volume = {45},
  issue = {1},
  pages = {304--307},
  numpages = {0},
  year = {1992},
  month = {Jan},
  publisher = {American Physical Society},
  doi = {10.1103/PhysRevB.45.304},
}

@article{SPT_digital_22,
	title = {Digital quantum simulation of {Floquet} symmetry-protected topological phases},
	volume = {607},
	issn = {0028-0836, 1476-4687},
	doi = {10.1038/s41586-022-04854-3},
	language = {en},
	number = {7919},
	urldate = {2026-07-15},
	journal = {Nature},
	author = {Zhang, Xu and Jiang, Wenjie and Deng, Jinfeng and Wang, Ke and Chen, Jiachen and Zhang, Pengfei and Ren, Wenhui and Dong, Hang and Xu, Shibo and Gao, Yu and Jin, Feitong and Zhu, Xuhao and Guo, Qiujiang and Li, Hekang and Song, Chao and Gorshkov, Alexey V. and Iadecola, Thomas and Liu, Fangli and Gong, Zhe-Xuan and Wang, Zhen and Deng, Dong-Ling and Wang, H.},
	month = jul,
	year = {2022},
	pages = {468--473},
}

@article{SPT_SZM_25,
	title = {Topological prethermal strong zero modes on superconducting processors},
	volume = {645},
	issn = {0028-0836, 1476-4687},
	doi = {10.1038/s41586-025-09476-z},
	language = {en},
	number = {8081},
	urldate = {2026-07-15},
	journal = {Nature},
	author = {Jin, Feitong and Jiang, Si and Zhu, Xuhao and Bao, Zehang and Shen, Fanhao and Wang, Ke and Zhu, Zitian and Xu, Shibo and Song, Zixuan and Chen, Jiachen and Tan, Ziqi and Wu, Yaozu and Zhang, Chuanyu and Gao, Yu and Wang, Ning and Zou, Yiren and Zhang, Aosai and Li, Tingting and Zhong, Jiarun and Cui, Zhengyi and Han, Yihang and He, Yiyang and Wang, Han and Yang, Jia-Nan and Wang, Yanzhe and Shen, Jiayuan and Liu, Gongyu and Deng, Jinfeng and Dong, Hang and Zhang, Pengfei and Li, Weikang and Yuan, Dong and Lu, Zhide and Sun, Zheng-Zhi and Li, Hekang and Zhang, Junxiang and Song, Chao and Wang, Zhen and Guo, Qiujiang and Machado, Francisco and Kemp, Jack and Iadecola, Thomas and Yao, Norman Y. and Wang, H. and Deng, Dong-Ling},
	month = sep,
	year = {2025},
	pages = {626--632},
}

@article{SPT_NRP_26,
	title = {Simulating topological order on quantum processors},
	volume = {8},
	issn = {2522-5820},
	doi = {10.1038/s42254-025-00911-8},
	language = {en},
	number = {3},
	urldate = {2026-07-15},
	journal = {Nat Rev Phys},
	author = {Gammon-Smith, Adam and Knap, Michael and Pollmann, Frank},
	month = jan,
	year = {2026},
	pages = {160--170},
}

@article{SPT_Sciece_25,
    author = {Haoran Qian  and Ming Gong  and Jiahui Zhang  and Shaojun Guo  and Chen Zha  and Fusheng Chen  and Yangsen Ye  and Yulin Wu  and Sirui Cao  and Chong Ying  and Qingling Zhu  and He-Liang Huang  and Youwei Zhao  and ShaoWei Li  and Jiale Yu  and Daojin Fan  and Dachao Wu  and Hong Su  and Hui Deng  and Hao Rong  and Yuan Li  and Kaili Zhang  and Tung-Hsun Chung  and Futian Liang  and Jin Lin  and Yu Xu  and Cheng Guo  and Na Li  and Kai Yan  and Fei-Fan Su  and Gang Wu  and Yong-Heng Huo  and Cheng-Zhi Peng  and Chao-Yang Lu  and Feng Mei  and Suotang Jia  and Xiaobo Zhu  and Jian-Wei Pan },
    title = {Programmable higher-order nonequilibrium topological phases on a superconducting quantum processor},
    journal = {Science},
    volume = {390},
    number = {6776},
    pages = {930-934},
    year = {2025},
    doi = {10.1126/science.adp6802},
}

@article{SPT_QC_prl20,
  title = {Symmetry-Enhanced Boundary Qubits at Infinite Temperature},
  author = {Kemp, Jack and Yao, Norman Y. and Laumann, Chris R.},
  journal = {Phys Rev Lett},
  volume = {125},
  issue = {20},
  pages = {200506},
  numpages = {6},
  year = {2020},
  month = {Nov},
  publisher = {American Physical Society},
  doi = {10.1103/PhysRevLett.125.200506},
}

\section*{Data availability}
The source data generated in this study are provided with this paper as Source Data files.
\end{document}

\title{Supplementary Materials for ``Rydberg-Atom-Mediated Strong Antisymmetric Spin Exchange in Molecular Arrays"}
%\title{Supplementary Information for ``Rydberg-atom-mediated strong ultracold molecular spin-exchange interactions in optical tweezer arrays"}

\author{Yunqing Jiao}
\thanks{These authors contributed equally to this work.}
\affiliation{State Key Laboratory of Quantum Optics Technologies and Devices, Institute of Laser Spectroscopy, Shanxi University, Taiyuan, Shanxi 030006, China}

\author{Jin-Zhu Jiang}
\thanks{These authors contributed equally to this work.}
\affiliation{State Key Laboratory of Quantum Optics Technologies and Devices, Institute of Laser Spectroscopy, Shanxi University, Taiyuan, Shanxi 030006, China}

\author{Bo-Wen Guan}
\affiliation{State Key Laboratory of Quantum Optics Technologies and Devices, Institute of Laser Spectroscopy, Shanxi University, Taiyuan, Shanxi 030006, China}

\author{Jie Ma}
\affiliation{State Key Laboratory of Quantum Optics Technologies and Devices, Institute of Laser Spectroscopy, Shanxi University, Taiyuan, Shanxi 030006, China}
\affiliation{Collaborative Innovation Center of Extreme Optics, Shanxi University, Taiyuan, Shanxi 030006, China}

\author{Liantuan Xiao}
\affiliation{State Key Laboratory of Quantum Optics Technologies and Devices, Institute
	of Laser Spectroscopy, Shanxi University, Taiyuan, Shanxi 030006, China}
\affiliation{Collaborative Innovation Center of Extreme Optics, Shanxi University, Taiyuan, Shanxi 030006, China}

\author{Chi Zhang}
\email{c.zhang@imperial.ac.uk}
\affiliation{Centre for Cold Matter, Blackett Laboratory, Imperial College London, Prince Consort Road, London SW7 2AZ, United Kingdom}

\author{Weibin Li}
\email{weibin.li@nottingham.ac.uk}
\affiliation{School of Physics and Astronomy, and Centre for the Mathematics and Theoretical Physics of Quantum Non-equilibrium Systems, The University of Nottingham, Nottingham NG7 2RD, United Kingdom}

\author{Feng Mei}
\email{meifeng@sxu.edu.cn}
\affiliation{State Key Laboratory of Quantum Optics Technologies and Devices, Institute of Laser Spectroscopy, Shanxi University, Taiyuan, Shanxi 030006, China}
\affiliation{Collaborative Innovation Center of Extreme Optics, Shanxi University, Taiyuan, Shanxi 030006, China}

\author{Suotang Jia}
\affiliation{State Key Laboratory of Quantum Optics Technologies and Devices, Institute
	of Laser Spectroscopy, Shanxi University, Taiyuan, Shanxi 030006, China}
\affiliation{Collaborative Innovation Center of Extreme Optics, Shanxi University, Taiyuan, Shanxi 030006, China}

\maketitle
\tableofcontents

\section*{Supplementary Note 1: Rydberg-atom-mediated effective molecular DM interaction Hamiltonian}
In this section, we provide a detailed derivation that connects the projected two-step evolution operator to the effective DM interaction Hamiltonian between the molecular spins. After restricting the two-step unitary operation $U_2(\alpha-\beta)$ [Eq.~(7) of the main text] to the $n_{\mathrm{exc}}=1$ subspace $\{\left|\downarrow R_1 \uparrow\right\rangle,\ \left|\uparrow R_1 \downarrow\right\rangle\}$, the evolution operator becomes
\begin{align}
U_{2}^\prime
&=
\begin{pmatrix}
\cos(\alpha-\beta) & -\sin(\alpha-\beta) \\
\sin(\alpha-\beta) & \cos(\alpha-\beta)
\end{pmatrix}.
\label{Umol}
\end{align}
This matrix is generated by a Pauli-$y$ operator acting in the ordered single-excitation basis $\{|\downarrow\uparrow\rangle,|\uparrow\downarrow\rangle\}$. Defining
\begin{equation}
S^y_{n_{\mathrm{exc}}=1}=-i|\downarrow\uparrow\rangle\langle\uparrow\downarrow|+i|\uparrow\downarrow\rangle\langle\downarrow\uparrow|,
\end{equation}
the projected unitary can be written as
\begin{equation}
U_2'=\exp[-i(\alpha-\beta)S^y_{n_{\mathrm{exc}}=1}]
=\exp(-iH_{\mathrm{DM}}^z t),
\end{equation}
where $t$ is the total time of the two-step protocol. The corresponding effective Hamiltonian is
\begin{equation}
\begin{aligned}
H_{\mathrm{DM}}^z
&=J_{\mathrm{eff}}S^y_{n_{\mathrm{exc}}=1}\\
&=J_{\mathrm{eff}}\left(-iS_1^-S_2^+ + iS_1^+S_2^-\right)\\
&=\frac{J_{\mathrm{eff}}}{2}
\left(S_1^xS_2^y-S_1^yS_2^x\right)\\
&=D_{12}^z(\mathbf S_1\times\mathbf S_2)_z,
\end{aligned}
\end{equation}
with $J_{\mathrm{eff}}=(\alpha-\beta)/{t}, D_{12}^z=J_{\mathrm{eff}}/{2}=(\alpha-\beta)/{2t}.$ An analogous derivation applies to the double-excitation manifold $\{\left|\downarrow R_2 \uparrow\right\rangle,\ \left|\uparrow R_2 \downarrow\right\rangle\}$, which undergoes a closed evolution under the same two-step protocol and results in an effective DM-$z$ interaction Hamiltonian.

\section*{Supplementary Note 2: Robustness of Rydberg-atom-mediated molecular gate operations}
In this section, we describe the model used to calculate the fidelities of Bell-state preparation and the SWAP and CNOT gates shown in Fig.~2(b,d,f) of the main text. The dynamics obey the Lindblad master equation
\begin{equation}
\dot{\rho}(t)
= -i[H,\rho(t)]
+ \sum_k \mathcal{D}[L_k]\rho(t),
\end{equation}
with the dissipator defined as
\begin{equation}
\mathcal{D}[L_k]\rho
= L_k\rho L_k^\dagger 
- \frac{1}{2}\{L_k^\dagger L_k,\,\rho\}.
\end{equation}
In the simulations, the master equation is solved within a truncated Hilbert space that explicitly contains the two molecular qubits and a minimal three-level subspace of the Rydberg atom, consisting of the two selected Rydberg states $\{|R_1\rangle,|R_2\rangle\}$ and the atomic ground state $|g\rangle$. The ground state $|g\rangle$ serves solely as a radiative decay endpoint and does not participate in any coherent dynamics, since the optical Rabi coupling $\Omega$ between $|g\rangle$ and the Rydberg manifold is turned off during the evolution. This restricted description faithfully captures both the spin-exchange dynamics and the dominant dissipative channel relevant on the timescale of these operations. Within this reduced space, the system is governed by the effective spin-exchange Hamiltonian of Eq.~(2) in the main text, supplemented by a finite detuning $\Delta$ between the molecular and Rydberg transition frequencies. Setting the molecular rotational transition angular frequency to $\omega=2\pi\times3471.8~\mathrm{MHz}$, the coherent Hamiltonian takes the form
\begin{equation}
H =
\frac{J_1}{2}\!\left(S_{1}^{+} S_{\mathrm{R}}^{-} + S_{1}^{-} S_{\mathrm{R}}^{+}\right)
+
\frac{J_2}{2}\!\left(S_{\mathrm{R}}^{+} S_{2}^{-} + S_{\mathrm{R}}^{-} S_{2}^{+}\right)
+
\frac{\omega}{2}\left(S_{1}^{z} + S_{2}^{z}\right)
+
\frac{\omega + \Delta}{2}\,S_{\mathrm{R}}^{z}.
\end{equation}
Dissipation arises from radiative decay of the two Rydberg states $|R_1\rangle = |nL_{j,m_J}\rangle = |72P_{3/2,3/2}\rangle$ and $|R_2\rangle = |71D_{5/2,5/2}\rangle$, modeled through the Lindblad operators
\begin{equation}
L_1=\sqrt{\Gamma_{R_1}}\,|g\rangle\!\langle R_1|, \quad 
L_2=\sqrt{\Gamma_{R_2}}\,|g\rangle\!\langle R_2|.
\end{equation}
Molecular dissipation is neglected because the coherence times of encoded rotational states are generally much longer than the microsecond-scale operation times considered here.

For each quasi-static noise realization, the detuning fluctuation $\Delta$ and atomic-position uncertainty $\Delta r$ are held fixed throughout the evolution. The motional drift of the untrapped Rydberg atom during the microsecond-scale evolution time is negligible (typically tens of nanometers), well below the few-percent distance variations explored in Fig.~2. The corresponding exchange amplitudes are  
\begin{equation}
J_{1}=\frac{\kappa\, d^{\uparrow\downarrow}_{1} d^{\uparrow\downarrow}_{\mathrm{R}}}{(r_{1}-\Delta r)^3},\quad
J_{2}=\frac{\kappa\, d^{\uparrow\downarrow}_{2} d^{\uparrow\downarrow}_{\mathrm{R}}}{(r_{2}+\Delta r)^3}.
\end{equation}
The system is initialized in the corresponding input state for each fidelity calculation. After the prescribed evolution time $t$, the reduced molecular state is obtained by tracing out the Rydberg ancilla,
\begin{equation}
	\rho_{\mathrm M}(t)
	=
	\operatorname{Tr}_{\mathrm R}
	\!\left[\rho(t)\right].
\end{equation}
The fidelity is then evaluated as
\begin{equation}
	F=
	\left[
	\operatorname{Tr}
	\sqrt{
		\sqrt{\rho_{\mathrm M}(t)}\,
		\rho_{\mathrm{target}}\,
		\sqrt{\rho_{\mathrm M}(t)}
	}
	\right]^2,
\end{equation}
where $\rho_{\mathrm{target}}$ denotes the ideal molecular target state for Bell-state preparation or the ideal output of the SWAP/CNOT gate. In calculating the CNOT fidelity, the single-qubit rotation gates $R_\mu(\theta)=\exp(-i\theta S^\mu/2),\mu\in\{x,y,z\}$ are treated as ideal instantaneous operations. The fidelity maps reported in Fig.~2(b,d,f) of the main text are then obtained by scanning over $(\Delta,\Delta r)$.

\section*{Supplementary Note 3: Floquet Hamiltonian engineering from $H_{\mathrm{DM}}^z$ to  $H_{\mathrm{DM}}^{xyz}$}
In this section, we show how the native DM-$z$ Hamiltonian can be generalized to a fully anisotropic DM-$xyz$ Hamiltonian through Floquet Hamiltonian engineering. We begin with a brief review of the average Hamiltonian theory. Including the control drive, the Hamiltonian reads
\begin{equation}
\begin{aligned}
&H(t) = H_{\mathrm{DM}}^{z} + H_{\mathrm{drive}}(t),\\
&H_{\mathrm{DM}}^{z} = D_{12}^z\left(S^{x}_{1} S^{y}_{2}-S^{y}_{1} S^{x}_{2}\right),\\
&H_{\mathrm{drive}}(t)=\sum_n \frac{\Omega(t)}{2}\left[\cos\phi(t)\,S^x_n+\sin\phi(t)\,S^y_n\right].
\end{aligned}
\end{equation}
The toggling-frame Hamiltonian is defined by
$$
U_{\mathrm{drive}}(t)=\mathcal{T}\exp\!\left[-i\!\int_0^t H_{\mathrm{drive}}(\tau)d\tau\right],\qquad 
\tilde{H}(t)=U_{\mathrm{drive}}^\dagger(t)\,H_{\mathrm{DM}}^z\,U_{\mathrm{drive}}(t).
$$
Over one Floquet period $T$, the stroboscopic evolution $U(T)=e^{-iH_{\mathrm{Floq}}T}$ is governed by the Magnus expansion $H_{\mathrm{Floq}}=H^{(0)}+H^{(1)}+\cdots$, where
$$
H^{(0)}
=\frac{1}{T}\int_0^T \tilde H(\tau)\,d\tau,
\qquad
H^{(1)}=-\frac{i}{2T}\int_0^T d\tau_2\int_0^{\tau_2} d\tau_1\,[\tilde H(\tau_2),\tilde H(\tau_1)].
$$
A time-reversal-symmetric sequence suppresses all odd Magnus terms and strongly reduces higher-order corrections, so that the effective Hamiltonian is well approximated by
\begin{equation}
H_{\mathrm{Floq}}\simeq H^{(0)}=\frac{1}{T}\sum_i \tilde{H}_i t_i.
\label{Heff}
    \end{equation}

\begin{figure}[t]
\centering
\includegraphics[scale=0.8]{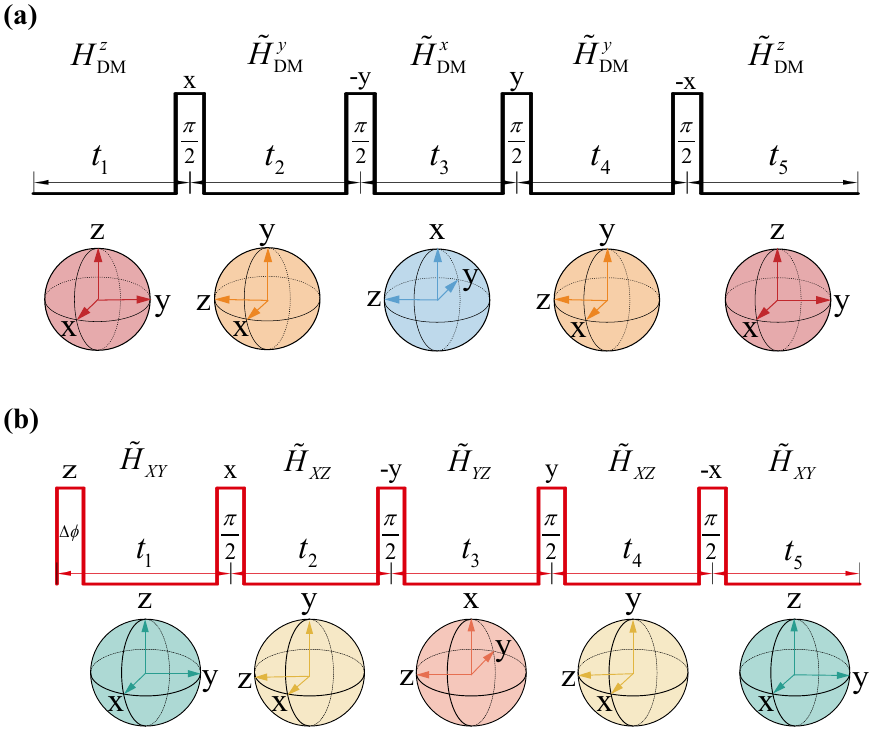}
	\caption{\textbf{Generalization of $H_{\mathrm{DM}}^z$ via Floquet Hamiltonian Engineering.}
    (a) Pulse sequence implementing a DM-$xyz$ Hamiltonian.  
	(b) Sequence generating an XYZ-type Hamiltonian.}
\label{FigS1}
\end{figure}
Within this framework, the five-segment pulse sequence in Fig.~\ref{FigS1}a and Table~\ref{DMXYZ} cyclically maps the native $S^xS^y-S^yS^x$ interaction onto the three components of the Dzyaloshinskii--Moriya vector. \begin{table}[htbp]
\centering
\begin{tabular}{c c c c c c c c c c c }
\hline
$H_{\mathrm{DM}}^z/D_{12}^z$& \hspace{0.5cm} & $t_1$ & $R_{x}(\pi/2)$ & $t_2$ & $R_{-y}(\pi/2)$ & $t_3$ & $R_{y}(\pi/2)$ & $t_4$ & $R_{-x}(\pi/2)$ & $t_5$ \\
\hline
$S^x_1 S^y_2$ &  & $\tilde{S}^x_1 \tilde{S}^y_2$ & $\rightarrow$ & $-\tilde{S}^x_1 \tilde{S}^z_2$& $\rightarrow$ & $\tilde{S}^y_1 \tilde{S}^z_2$ & $\rightarrow$ & $-\tilde{S}^x_1 \tilde{S}^z_2$ & $\rightarrow$ & $\tilde{S}^x_1 \tilde{S}^y_2$ \\
- & &  & & & & & & & & \\
$S^y_1 S^x_2$ & & $\tilde{S}^y_1 \tilde{S}^x_2$ & $\rightarrow$ & $-\tilde{S}^z_1 \tilde{S}^x_2$ & $\rightarrow$ & $\tilde{S}^z_1 \tilde{S}^y_2$ & $\rightarrow$ & $-\tilde{S}^z_1 \tilde{S}^x_2$ & $\rightarrow$ & $\tilde{S}^y_1 \tilde{S}^x_2$ \\
\hline
\end{tabular}
\caption{Floquet  pulse sequence transforming $H_{\mathrm{DM}}^{z}$ into $H_{\mathrm{DM}}^{xyz}$.}
\label{DMXYZ}
\end{table}Each component accumulates during the corresponding free-evolution segments, with its effective coupling determined by their total duration. 
Substituting the toggled-frame Hamiltonians into Eq.~\eqref{Heff} gives
\begin{equation}
H_{\mathrm{DM}}^{xyz}
=D_Z(S_1^xS_2^y-S_1^yS_2^x)
+D_Y(S_1^zS_2^x-S_1^xS_2^z)
+D_X(S_1^yS_2^z-S_1^zS_2^y),
\end{equation}
which is an anisotropic DM-$xyz$ interaction. Here
$
D_Z=D_{12}^z\frac{t_1+t_5}{T},\ 
D_Y=D_{12}^z\frac{t_2+t_4}{T},\ 
D_X=D_{12}^z\frac{t_3}{T};\   
t_1=t_5,\ t_2=t_4,\ T=\sum_i t_i.
$

\section*{Supplementary Note 4: Floquet Hamiltonian engineering from $H_{\mathrm{DM}}^z$ to $H_{\mathrm{XYZ}}$}
Building on the general framework of average Hamiltonian theory introduced in Supplementary Note 3, we now apply the same Floquet-engineering strategy to transform the native DM-$z$ Hamiltonian into an XYZ Hamiltonian. Tunable XYZ-type interactions provide a versatile platform for exploring non-equilibrium quantum dynamics in interacting many-body spin models~\cite{DDI_gorshkovPRL_11,DDI_gorshkovPRA_11,DDI_JYeSci_25,DDI_JYeNat_24}. Specifically, we seek to engineer a Hamiltonian of the form $H_{\mathrm{XYZ}}=J_\mathrm{X}S_1^xS_2^x+J_\mathrm{Y}S_1^yS_2^y+J_\mathrm{Z}S_1^zS_2^z$. A relative single-qubit phase rotation,
$U^z=e^{\frac{i}{2}(S_1^z\phi_1+S_2^z\phi_2)},$
is first applied to convert the DM-$z$ coupling into a controllable superposition of the DM-$z$ and XY operators. With the relative phase defined as $\Delta\phi=\phi_1-\phi_2$, the transformed Hamiltonian becomes
\begin{equation}
H_{\mathrm{eff}}
= D_{12}^z\!\left[\cos(\Delta\phi)\left(S_1^x S_2^y-S_1^y S_2^x\right)
+\sin(\Delta\phi)\left(S_1^x S_2^x+S_1^y S_2^y\right)\right].
\end{equation}
Setting $\Delta\phi=\pi/2$ isolates the XY interaction,
$
\tilde H_{\mathrm{XY}}=D_{12}^z(\tilde S_1^x\tilde S_2^x+\tilde S_1^y\tilde S_2^y),
$
which forms the starting point of the toggling sequence in Table~\ref{XYZ}. Subsequent $\pi/2$ rotations cycle the interaction through the XY, XZ, and YZ forms, with the corresponding free-evolution times determining the final XYZ couplings (Fig.~\ref{FigS1}b).
\begin{table}[htbp]
\centering
\begin{tabular}{c c c c c c c c c c c c c }
\hline
$H_{\mathrm{DM}}^z/D_{12}^z$ &$\Delta\phi_{-z}$& $\tilde H_{\mathrm{XY}}/D_{12}^z$&\hspace{0.5cm} 
& $t_1$ &$R_{x}(\pi/2)$ & $t_2$ & $R_{-y}(\pi/2)$ & $t_3$ & $R_{y}(\pi/2)$ & $t_4$ & $R_{-x}(\pi/2)$ & $t_5$ \\
\hline
$S^x_1 S^y_2$ & $\rightarrow$ & $\tilde{S}^x_1 \tilde{S}^x_2$&
& $\tilde{S}^x_1 \tilde{S}^x_2$  & $\rightarrow$ &
$\tilde{S}^x_1 \tilde{S}^x_2$ & $\rightarrow$ &
$\tilde{S}^y_1 \tilde{S}^y_2$ & $\rightarrow$ &
$\tilde{S}^x_1 \tilde{S}^x_2$ & $\rightarrow$ &
$\tilde{S}^x_1 \tilde{S}^x_2$\\
- & & + & & & & & & & & & & \\
$S^y_1 S^x_2$ & $\rightarrow$ & $\tilde{S}^y_1 \tilde{S}^y_2$&
& $\tilde{S}^y_1 \tilde{S}^y_2$ & $\rightarrow$ &
$\tilde{S}^z_1 \tilde{S}^z_2$ & $\rightarrow$ &
$\tilde{S}^z_1 \tilde{S}^z_2$ & $\rightarrow$ &
$\tilde{S}^z_1 \tilde{S}^z_2$ & $\rightarrow$ &
$\tilde{S}^y_1 \tilde{S}^y_2$ \\
\hline
\end{tabular}
\caption{Floquet pulse sequence transforming $H_{\mathrm{DM}}^{z}$ into $H_{\mathrm{XYZ}}$.}
\label{XYZ}
\end{table}
The resulting effective Hamiltonian reads
\begin{equation}
H_{\mathrm{XYZ}}=J_\mathrm{X}S_1^xS_2^x+J_\mathrm{Y}S_1^yS_2^y+J_\mathrm{Z}S_1^zS_2^z,
\end{equation}
with coefficients controlled by the segment durations 
$
J_\mathrm{X} = D_{12}^z\frac{t_1+t_2+t_4+t_5}{T},\ 
J_\mathrm{Y} = D_{12}^z\frac{t_1+t_3+t_5}{T},\ 
J_\mathrm{Z} = D_{12}^z\frac{t_2+t_3+t_4}{T},\   
t_1=t_5,\ t_2=t_4,\ T=\sum_i t_i.
$
When the accumulated evolution times satisfy $t_1+t_5=t_2+t_4=t_3$, the three couplings become equal, $J_\mathrm{X}=J_\mathrm{Y}=J_\mathrm{Z}=2D_{12}^z/3$, and the effective Hamiltonian reduces to the isotropic Heisenberg form.

\section*{Supplementary Note 5: Long-lived edge coherence in the non-equilibrium SPT phase}\label{FSPT}
In this section, we analyze the exactly solvable point $DT=1.5\pi$ to elucidate the origin of the long-lived edge coherence characteristic of the non-equilibrium SPT phase. At this point, the edge spins decouple exactly from the bulk, allowing the underlying edge dynamics to be established analytically. We consider the three-step periodic drive used in the main text on an open chain of $N$ qubits:
\begin{equation}
	H(t)=
	\begin{cases}
    	\displaystyle \ \sum_{n} \ \ g_{n}\,S^x_{n}, & 0\le t<T/3, \\[6pt]
    	\displaystyle \sum_{n\in\mathrm{odd}} D\,\big(S^x_{n}S^y_{n+1}-S^y_{n}S^x_{n+1}\big), & T/3\le t<2T/3, \\[6pt]
    	\displaystyle \sum_{n\in\mathrm{even}} D\,\big(S^x_{n}S^y_{n+1}-S^y_{n}S^x_{n+1}\big), & 2T/3\le t<T.
	\end{cases}
\end{equation}
Here $D\equiv D^z_{n,n+1}$ is the nearest-neighbor DM interaction strength. We denote the unitary generated by the disordered onsite-rotation step as $U_x$, and those generated by the odd- and even-bond DM interaction steps as $U_o$ and $U_e$, respectively. The one-period evolution is then $U(T)=U_eU_oU_x$. At the fixed point $DT=1.5\pi$, each DM step accumulates an interaction angle $DT/3=0.5\pi$. The two operators $S_n^xS_{n+1}^y$ and $S_n^yS_{n+1}^x$ on a single bond satisfy
$(S_n^xS_{n+1}^y)^2=(S_n^yS_{n+1}^x)^2=\mathbb I$ and $[S_n^xS_{n+1}^y,S_n^yS_{n+1}^x]=0$. Therefore,
\begin{equation}
    \begin{aligned}
        \exp\left[-i\frac{\pi}{2}(S_n^xS_{n+1}^y-S_n^yS_{n+1}^x)\right]
        &=\exp\left(-i\frac{\pi}{2}S_n^xS_{n+1}^y\right)\exp\left(+i\frac{\pi}{2}S_n^yS_{n+1}^x\right)\\
        &=(-iS_n^xS_{n+1}^y)(iS_n^yS_{n+1}^x)\\
        &=S_n^zS_{n+1}^z.
    \end{aligned}
\end{equation}
Consequently, each DM bond operation reduces to a nearest-neighbor $S^zS^z$ operator. Combining the odd- and even-bond layers gives:
\begin{equation}
U_{\mathrm{e}}U_{\mathrm{o}}=\prod_{n=1}^{N-1} S_n^zS_{n+1}^z=S_1^zS_N^z,
\end{equation}
because every bulk Pauli operator appears twice and squares to the identity. Including the onsite-rotation step then gives
\begin{equation}
U(T)=
S_1^zS_N^z
\exp\left(-i\frac{T}{3}\sum_{n=1}^{N}g_nS_n^x\right).
\end{equation}
Thus, $U(T)$ factorizes into an edge-only operator $S^z_1S^z_N$ and the product of single-qubit $x$ rotations generated during the first step. This exact result at the solvable point makes explicit the dynamical decoupling between the edge and bulk degrees of freedom, providing an analytical picture of the long-lived edge coherence characteristic of the non-equilibrium SPT phase.

\bibliographystyle{sb-vancouver3}
\bibliography{refv4}